\documentclass[11pt,a4paper]{article}
\pdfoutput=1
\usepackage{jheppub}
\usepackage{amsmath,amscd}
\usepackage{amsfonts}
\usepackage{amssymb}
\usepackage{graphicx}
\usepackage{color}
\usepackage[normalem]{ulem}

\def\Re{{\rm Re}}
\def\Im{{\rm Im}}

\newcommand{\RR}{\mathbb{R}} 
\newcommand{\ZZ}{\mathbb{Z}} 
\newcommand{\NN}{\mathbb{N}} 

\def\cala         {{\cal A}}

\def\cald         {{\cal D}}

\def\calg         {{\cal G}}

\def\calk         {{\cal K}}
\def\call         {{\cal L}}
\def\calm         {{\cal M}}
\def\caln         {{\cal N}}
\def\calo         {{\cal O}}
\def\calp         {{\cal P}}

\def\calr         {{\cal R}}

\def\calt         {{\cal T}}

\newsavebox{\uuunit}
\sbox{\uuunit}
    {\setlength{\unitlength}{0.825em}
     \begin{picture}(0.6,0.7)
        \thinlines
        \put(0,0){\line(1,0){0.5}}
        \put(0.15,0){\line(0,1){0.7}}
        \put(0.35,0){\line(0,1){0.8}}
       \multiput(0.3,0.8)(-0.04,-0.02){12}{\rule{0.5pt}{0.5pt}}
     \end {picture}}

\def\be{\begin{equation}}
\def\ee{\end{equation}}
\def\bea{\begin{eqnarray}}
\def\eea{\end{eqnarray}}

\def\a{\alpha}

\def\h{\eta}
\def\g{\gamma}

\def\d{\delta}
\def\e{\epsilon}
\def\D{\Delta}
\def\l{\lambda}

\def\m{\mu}
\def\n{\nu}

\def\p{\pi}
\def\r{\rho}

\def\s{\sigma}

\def\t{\tau}
\def\x{\xi}

\def\bh{{\bar{h}}}

\def\sF{{{ F}\!\!\!\!\hskip.8pt\hbox{\raise1pt\hbox{/}}\,}}
\def\som{{{ \omega}\!\!\!\!\hskip.8pt\hbox{\raise1pt\hbox{/}}\,}}
\def\sJ{{{\rm J}\!\!\!\!\hskip.8pt\hbox{\raise1pt\hbox{/}}\,}}

\def\F{\Phi}
\def\pa{\partial}

\def\to{\rightarrow}
\def\nonu{\nonumber \\{}}
\def\half{{1 \over 2}}

\title{Chiral boundary conditions for singletons and W-branes}
\author[a]{Joris Raeymaekers,}
\author[b,c]{Dieter Van den Bleeken}

\affiliation[a]{Institute of Physics of the ASCR, \\
Na Slovance 2, 182 21 Prague 8, Czech Republic.}
\affiliation[b]{Primary address\\
 Physics Department, Bo\u{g}azi\c{c}i University\\
 34342 Bebek / Istanbul, Turkey}
\affiliation[c]{Secondary address\\
Institute for Theoretical Physics, KU Leuven\\
3001 Leuven, Belgium}

\emailAdd{joris@fzu.cz}
\emailAdd{dieter.van@boun.edu.tr}

\abstract{We revisit the holographic dictionary for a free massless scalar in AdS$_3$, focusing on the `singleton' solutions for which the boundary profile is an arbitrary chiral function.  We look for consistent boundary conditions which include this class of solutions. On one hand, we give a no-go argument that they cannot be interpreted  within any boundary condition
which preserves full conformal invariance. On the other hand, we show that such solutions fit naturally in a generalization of the Comp\`{e}re-Song-Strominger boundary conditions, which preserve a chiral Virasoro and current algebra. These observations have implications for the black hole deconstruction proposal, which proposes singleton solutions as candidate black hole microstate geometries.
Our results suggest that the chiral boundary condition, which also contains the extremal BTZ black hole, is the natural setting for holographically interpreting the black hole deconstruction proposal. }

\begin{document}
\maketitle

\section{Introduction}
In this work we revisit the boundary conditions and holography for   a massless complex scalar field in AdS$_3$.
Concretely, we will be  interested in the holographic interpretation of  solutions of the  wave equation $\Box t =0$ which are given by an arbitrary holomorphic function of a complex variable $v$:
\begin{equation}
t=g(v)\,,\qquad\qquad v= e^{-ix_-}\tanh \r\,.\label{singletonsols}
\end{equation}
 Here, we work in global coordinates in
terms of which the AdS$_3$ metric is
\be
ds^2 = l^2 \left[-\cosh^2 \r dT^2 +  d\r^2 + \sinh^2 \r d\varphi^2\right]\label{globalcoord}
\ee
and we set $x_\pm = T\pm \varphi$.
 From (\ref{singletonsols})  we can of course generate other classes of solutions depending on a free function by using discrete symmetries such as parity, $x_- \leftrightarrow x_+$, or charge conjugation, $t \leftrightarrow \bar t$.
We will focus on the set (\ref{singletonsols}), since they can be shown to preserve supersymmetry in parity-breaking theories where  supersymmetry resides in the `left-moving' sector.

Expanding the solutions  (\ref{singletonsols}) near the AdS$_3$ boundary, {one sees} that they don't {vanish} there, but {asymptote} to an arbitrary right-moving function $g(e^{-i x_-})$. This behaviour
is reminiscent of what happens in the `alternate' quantization, where the leading part of the scalar is allowed to fluctuate. {The case at hand is subtle} however, since the
 alternate quantization {is only generically defined} for massive scalars in the range \cite{Klebanov:1999tb}
 \be -1 < m^2 l^2 <0,\label{window}\ee
 where it corresponds to {assigning an}  operator  {of} dimension $\D_- = 1 - \sqrt{ 1 + m^2 l^2}$ {as the holographic dual}.
The massless scalar of interest  lies at the upper boundary of the window (\ref{window}) and requires special care, since
the would-be dual operator dimension   $\D_-=0$ then saturates the unitarity bound.
One might expect that in this case the field propagates a `short' multiplet. Indeed, as we will review below,  the solutions of interest (\ref{singletonsols}) carry a three-dimensional singleton \cite{Flato:1990eu} representation, which is  {a small} part of a larger representation which is reducible but not decomposable.

Our goal will therefore be to verify $a)$ whether a version of the `alternate' boundary condition can still be consistently\footnote{With consistent, we mean such that it leads to a consistent variational principle, i.e. the action functional is differentiable.} imposed so as to allow the
solutions of the type (\ref{singletonsols}) and $b)$ if so, if it preserves conformal symmetry. While the answer to  $a)$  will be affirmative, we will present a simple argument that
the answer to $b)$ is negative: there {exists an obstruction to} consistent boundary conditions which allow the solutions  (\ref{singletonsols}) while preserving the
global  $\mathrm{SL}(2,\mathbb{R})\times \overline{\mathrm{SL}(2,\mathbb{R})}$ symmetry of AdS$_3$. Interestingly, the boundary condition we propose in our answer to $a)$ does
preserve a purely right-moving Virasoro symmetry, combined with a right-moving U(1) current algebra: they are a natural generalization of the boundary conditions  proposed
by Comp\`{e}re, Song and Strominger \cite{Compere:2013bya} in the context of pure gravity.

While this generalization of  \cite{Compere:2013bya} to include matter is of interest in its own right, our main motivation for studying this question came from black hole physics, namely from a puzzle in
the  holographic interpretation of the black hole deconstruction proposal of Denef et. al. \cite{Denef:2007yt}. The idea of this work was to deconstruct   stringy black holes in terms
 bound states of zero-entropy D-brane centers, with a large moduli space to account for the black hole entropy.  A concrete realization for the  IIA D4-D0 black hole of \cite{Maldacena:1997de}
 was proposed to involve certain D2-brane configurations enveloping a D6-anti-D6 pair. The {D2's}  experience a magnetic field in the internal space and  their   large lowest Landau level degeneracy was {shown} to account for the entropy of this particular black hole. Upon taking an M-theory decoupling limit, these D2-brane configurations become particle-like objects
 in AdS$_3$, which source a complex scalar field.
They can  be seen as simple examples of so-called W-branes, the M-theory lift of open strings connecting D6-brane centers, which were conjectured to capture the entropy of a class of D-brane systems\footnote{Some puzzles with  the W-brane idea were pointed out in \cite{Tyukov:2016cbz}.} \cite{Martinec:2014gka},\cite{Martinec:2015pfa}.

  As was argued in \cite{Levi:2009az},\cite{Raeymaekers:2014bqa},\cite{Raeymaekers:2015sba}, the scalar profile in the black hole deconstruction solutions   is precisely of the form
 (\ref{singletonsols}). Therefore, our results can be interpreted as a no-go argument for the holographic interpretation of the  black hole deconstruction solutions within a standard CFT possessing two   Virasoro copies (such as the MSW CFT of \cite{Maldacena:1997de}). On the other hand, these solutions do belong to a dual chiral  theory with a right-moving Virasoro and current algebra. As we shall see, the extremal BTZ black {hole} that {one tries} to deconstruct also fits within the same boundary conditions, suggesting that this chiral theory{, which might be a deformation of the MSW CFT}, {should be} the proper setting for {a holographic interpretation of} the black hole deconstruction proposal.

This paper is structured as follows. In section \ref{secfree}, we revisit the classification of solutions to the massless wave equation and their representation content. We also
discuss boundary conditions and derive
a criterion to check whether a given  boundary condition  follows from a consistent variational principle. In section \ref{secnogo} we present our no-go argument for  interpreting the
singleton modes of the scalar within conformally invariant boundary conditions, and in section \ref{secchiral} we propose chiral boundary conditions which do include those solutions.
 In section \ref{secparticles} we consider examples: after the warm-up example of multi-centered solutions in pure gravity we turn to the
W-brane  solutions which also source a complex scalar. In the Discussion we focus on the implications of our results for the holographic interpretation of
the black hole deconstruction proposal.

\section{Revisiting the free massless scalar on AdS$_3$}\label{secfree}
In this section we revisit some of the physics of a free, massless, complex  scalar on a fixed AdS$_3$ background.
Our main motivation is the  application to the black hole deconstruction proposal in the following sections, but as this analysis might be of independent interest  we have tried to make this section self-contained.

We will review in some detail the classification of bulk solutions and their transformation under the action of the $\mathrm{SL}(2,\mathbb{R})\times \overline{\mathrm{SL}(2,\mathbb{R})}$ global symmetry of AdS$_3$, slightly expanding  on the results of \cite{Balasubramanian:1998sn} (see also \cite{Flato:1990eu},\cite{Starinets:1998dt}). We then go on to discuss in general the compatibility of a consistent variational principle with finite on-shell action and a choice of boundary conditions for this free scalar theory, reviewing the example of the standard Dirichlet boundary conditions. In the next two sections we
will focus on boundary conditions which allow for the solutions (\ref{singletonsols}) and their representation-theoretic content.


\subsection{Classification of solutions}
We consider a free, massless complex scalar on a fixed  AdS$_3$ background  with action\footnote{In the remainder of the paper we will work in units where $\mathrm{G}_\mathrm{N}=\frac{1}{16\pi}$}:
\be\label{SFree}
S = -{1\over 32\p \mathrm{G}_\mathrm{N}}\int d^3 x\, \sqrt{- G}\, {\pa_\m t \pa^\m \bar t}.
\ee
The equation of motion {obtained by} varying $\bar t$ is simply
\be
\Box t=0 \label{eomt}\,.
\ee
{It is convenient to introduce} Fefferman-Graham coordinates $(y,x_+,x_-)$, with $y = 4 e^{-2 \r}$, in terms of which the global AdS$_3$ metric takes the form
\be
\frac{ds^2_3}{l^2} = {dy^2 \over 4 y^2} -{1 \over y}\left(dx_+dx_-+\frac{y}{4}(dx_+^2+dx_-^2)+\frac{y^2}{16}dx_+dx_-\right)\label{ads3}
\ee
The above metric is invariant under an SL$(2,\mathbb{R})\times \overline{\mathrm{SL}(2,\mathbb{R})}$ isometry group generated by the Killing vectors
 \begin{align}
L_0 =&  i \pa_+, &
L_{\pm 1} = & i e^{\pm  i x_+} \left( \frac{16+y^2}{16-y^2} \pa_+ - {8y\over 16-y^2} \pa_- \pm i y\pa_y \right) \\
  \bar L_0 =&  i \pa_-,&
  \bar L_{\pm 1} = & i e^{\pm i x_-} \left( \frac{16+y^2}{16-y^2} \pa_- -{8y\over 16-y^2}  \pa_+ \pm i y \pa_y \right)
\end{align}
that {form the algebra}
\begin{align} [L_m, L_n] =& (m-n)L_{m+n} &  [\bar L_m,\bar L_n] =& (m-n)\bar L_{m+n}\label{Lalgebra}\end{align}

These isometries can be conveniently used to classify the solutions to the wave equation \eqref{eomt} as {the Laplacian} is a Casimir of the algebra above:
\begin{equation}
\Box=-\frac{2}{l^2}(L^2+\bar L^2)\qquad\qquad L^2=\frac{1}{2}\left(L_{-1}L_1+L_1L_{-1}\right)-L_0^2
\end{equation}
In particular {it follows} that solutions to this equation will form representations of the symmetry algebra with vanishing values for both $L^2$ and $\bar L^2$. Furthermore as $L_0$ and $\bar L_0$ commute with the Laplacian, we can label different solutions by their eigenvalues $h$ and $\bh$ under these operators, leading to solutions of the form
\begin{equation}
t_{h,\bh}(x_+,x_-,y)=e^{-i(h x_++\bh x_-)}f_{h,\bh}(y)\label{sepvar}
\end{equation}
Using this separation of variables \eqref{eomt} can be solved exactly in terms of hypergeometric functions. We give some details in appendix \ref{appmodesols}, but it is simpler and also sufficient for our purposes to do a  near-boundary analysis. Note that we will consider general solutions to \eqref{eomt}, without imposing any restrictions such as boundary- or regularity conditions for the moment.
The near-boundary behaviour of a general solution to (\ref{eomt}) is\footnote{The last, `logarithmic', term is special to the massless case  and would be  absent for a scalar with a mass in the window $-1 < m^2 l^2 <0$.} \cite{de Haro:2000xn}
\be
t = t_0 (x_+, x_-) + y\, t_2 (x_+, x_-)  +y\log y\, \tilde t_2  (x_+, x_-) + \calo (y^2\log y ).\label{tnb}
\ee
The equation of motion near the boundary leaves $t_0$ and $t_2$ arbitrary, while $\tilde t_2 $ is related to $t_0$ as
\be
\tilde t_2  = \pa_+ \pa_- t_0.
\ee
Combining this information with the separation of variables \eqref{sepvar}
one finds that for  given $h,\bh$ there are two types of solutions: the class $t^+_{h,\bh}$ for which the leading $t_0$ part in (\ref{tnb}) vanishes, and the class  $t^-_{h,\bh}$
for which it doesn't. Their near-boundary expansion is
\begin{eqnarray}
t^+_{h,\bh}&=&e^{-i(hx_++\bh x_-)}y+\calo(y^2\log y)\nonu
t^-_{h,\bh}&=&e^{-i(h x_++\bh x_-)}\left(1-h\bh y \left[\frac{1}{2|h|}+\psi(|h|)+\frac{1}{2|\bh|}+\psi(|\bh|)+\log y
\right]\right)+\calo(y^2\log y)\qquad |h|,|\bh|>0\nonu
t^-_{0,\bh;\sigma}&=&e^{-i\bh x_-}\left(1+\frac{1}{2}\sigma\bh y\right)+\calo(y^2\log y)\nonu
t^-_{h,0;\sigma}&=&e^{-ih x_+}\left(1+\frac{1}{2}\sigma hy\right)+\calo(y^2\log y)\label{basisas}
\end{eqnarray}
Note that we introduced an arbitrary parameter $\sigma$, which can be set to any preferred value by adding a certain multiple of $t^+_{0,\bh}$ or $t^+_{h,0}$ respectively. In addition to the
solutions discussed so far, there are also the zero-mode solutions which are not eigenstates of both $L_0$ and $\bar L_0$, namely
\be
t = x_+ \qquad {\rm and} \qquad t = x_-.\label{zeromodes}
\ee
Like the $t^-$ solutions, these also don't {vanish} near the boundary.

\subsection{Representation theory}
As discussed above, the set of solutions to the wave equation is guaranteed to carry a representation of the AdS symmetry group. It is well-known that this representation is neither unitary nor irreducible. In fact, the full set of scalar solutions carries a reducible but nondecomposable representation as we will now discuss in detail.

Before we begin, let us recall, and introduce some notation for, the unitary irreducible representations of  $\mathrm{SL}(2,\mathbb{R})$, see \cite{barut},\cite{Balasubramanian:1998sn} for more details. We will mainly restrict attention to the primary (lowest weight) and anti-primary (highest weight) representations.  We will use a simplified notation where both are denoted by  $(h)$, with the understanding that for  $h$  positive, we mean the primary representation built on a state with  $L_0$-eigenvalue $h$ and annihilated by $L_1$, while for $h$ negative we mean the anti-primary representation built on a state with  $L_0$-eigenvalue $h$ and annihilated by $L_{-1}$. The trivial representation will be denoted\footnote{In the  notation of \cite{Balasubramanian:1998sn}, $(h)$ with $h>0$ corresponds to $\cald^+(h,h)$, $(h)$ with $h<0$ to $\cald^-(h,-h)$ and $(0)$ to $\cald(0)$.} by $(0)$.
 The corresponding representations of the product group SL$(2,\mathbb{R})\times \overline{\mathrm{SL}(2,\mathbb{R})}$ will be denoted as $(h, \bar h)$.
The three-dimensional singleton representations of \cite{Flato:1990eu}   are   those  which are trivial with respect to one of the factors, i.e. of the type $(h,0)$ or  $(0,\bh)$. These representations are annihilated by $\bar L_{-1}$ resp. $L_{-1}$ and therefore short compared to the generic $(h,\bh )$ representation\footnote{We should alert the reader  that a different definition  of a 3D singleton was given in \cite{Gunaydin:1986fe}, namely as a representation carried by the Fock space of a single harmonic oscillator. In 4D both definitions of the singleton agree, but in 3D the latter definition gives the
  representations carried by a massive Klein-Gordon field with $m^2 l^2 = - 3/4$, which are not short in the sense described here.}.

Returning to the massless scalar solutions, the particular choice of basis \eqref{basisas}  is useful as it brings the representation of the algebra \eqref{Lalgebra} into a rather simple and disentangled form:
\begin{eqnarray}
L_0t_{h,\bh}^\pm=ht_{h,\bh}^\pm&\qquad& \bar L_0 t_{h,\bh}^\pm=\bh t_{h,\bh}^\pm\nonu
L_{\pm 1} t_{h,\bh}^+=(h\mp 1)t_{h\mp1,\bh}^+&\qquad& \bar L_{\pm 1} t_{h,\bh}^+=(\bh\mp 1)t_{h,\bh\mp1}^+\label{rep1}
\end{eqnarray}
\vspace{-0.9cm}
\begin{eqnarray}
L_{\pm 1} t_{h,\bh}^-=h t_{h\mp1,\bh}^-&\quad& \mbox{ when }h\neq 0,\pm 1\nonu
\bar L_{\pm 1} t_{h,\bh}^-=\bh t_{h,\bh\mp1}^-&\quad& \mbox{ when }\bh\neq 0,\pm 1\nonumber
\end{eqnarray}
\vspace{-0.9cm}
\begin{eqnarray}
L_{\mp 1} t_{\pm 1,\bh}^-=\pm t_{\pm 2,\bh}^-&\qquad&\bar L_{\mp 1} t_{h,\pm 1}^-=\pm t_{h,\pm2}^-\nonu
L_{\pm 1} t_{\pm 1,\bh}^-=\pm t_{0,\bh;\pm 1}^-&\qquad&\bar L_{\pm 1} t_{h,\pm 1}^-=\pm t_{h,0;\pm1}^-\nonu
L_{\pm 1} t_{0,\bh;\mp1}^-=0&\qquad&\bar L_{\pm 1} t_{h,0;\mp1}^-=0\nonu
L_{\pm 1} t_{0,\bh;\pm1}^-=-\bh t_{\mp 1,\bh}^+&\qquad&\bar L_{\pm 1} t_{h,0;\pm1}^-=-ht_{h,\mp 1}^+\label{rep2}
\end{eqnarray}

Imagine making a choice $h=h_0$ and $\bh=\bh_0$ and taking this as a starting point for acting on a corresponding solution repeatedly with any of the generators \eqref{Lalgebra}. In case $h_0$ and $\bh_0$ are non-integer the $t^+$ and $t^-$ fall in separate representations that contain all $h$ and $\bh$ such that $h-h_0, \bh-\bh_0\in\mathbb{Z}$. However when one (or both) of $h_0, \bh_0$ is integer things are somewhat different. Note that this includes the solutions of interest (\ref{singletonsols}), so we will focus on this case in what follows.  In particular if one starts with $t^-_{h_0,\bh_0}$ one will at some point end up at a state of the form $t^-_{0,\bh;\pm}$ or $t^-_{h,0;\pm}$, which then via the last line of the rules above feeds into states of the form $t^+$, meaning that we get a huge representation containing both $t^{+}$ and $t^-$ solutions. Note however that this argument passed through a one way street: one cannot produce $t^-$ states from $t^+$ states by acting with the generators.  Therefore
  the $t^+$ states form a subrepresentation on their own.

\subsubsection{Dirichlet modes with integer weights} \label{parDmodes} Let us describe in more detail the representation carried by the  $t^+$ modes when both $h$ and $\bh$ are integer. These can all be obtained from acting with the generators on the solution
\be
t^+_{0,0}=  2\ln {4+y \over 4-y}
\ee
Acting with $L_{-1}$ gives $t^+_{1,0}$, but acting with $L_1$ on $t^+_{1,0}$ gives zero instead of bringing us back to $t^+_{0,0}$. Therefore acting with the generators on $t^+_{1,0}$ we obtain an invariant subspace, whose complement is not invariant.  The representation is therefore reducible but nondecomposable.
 \begin{figure}
\begin{center}
\includegraphics[scale=0.6]{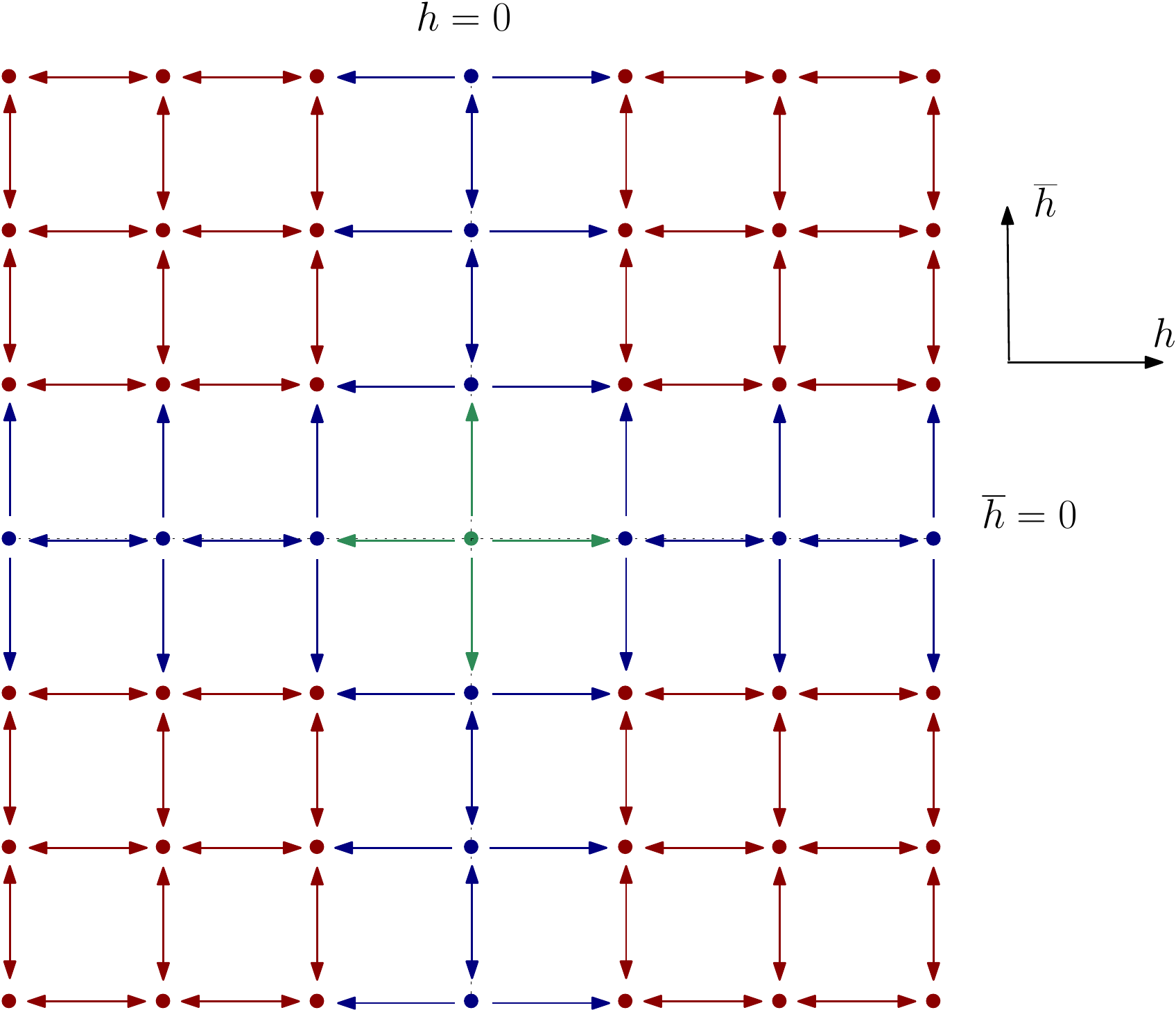}
\end{center}
\caption{A diagram showing the $\mathrm{SL}(2,\mathbb{R})\times \overline{\mathrm{SL}(2,\mathbb{R})}$ action on the $t^+_{h,\bh}$  solutions for integer $h$ and $\bh$.  Left/right arrows show non-trivial $L_\pm$ actions and down/up arrows indicate non-trivial $\bar L_\pm$ actions. The states in dark red form four unitary subrepresentations built on (anti-) primary states.  The total representation is indecomposable as there are irreversible lowering/raising actions starting from states with either $h$ or $\bh$ vanishing,  as can be seen from the presence of one-way arrows.}\label{tplusfig}
\end{figure}
Similar remarks apply  when acting on $t^+_{0,0}$ with the generators $L_{1}, \bar L_{\pm 1}$. The full structure of the representation is illustrated in figure \ref{tplusfig}, and can be schematically denoted in terms of irreducible representations as
\be\begin{array}{ccccc}
{\color{red}(-1,1)} & \leftarrow & {\color{blue} (0,1)} & \rightarrow & {\color{red}(1,1)}\\
\uparrow& &\uparrow& &\uparrow\\
{\color{blue}(-1,0)} & \leftarrow & {\color{green} (0,0)} & \rightarrow & {\color{blue}(1,0)}\\
\downarrow&&\downarrow&&\downarrow\\
{\color{red}(-1,-1)} & \leftarrow & {\color{blue} (0,-1)} & \rightarrow & {\color{red}(1,-1)}
\end{array}\label{tplustable}.\ee
We should note that the states in blue and green are often not considered as they are singular in the center of AdS$_3$.

\subsubsection{Singleton modes}\label{parsingmodes} Next we would like to describe the representation which contains the solutions of interest (\ref{singletonsols}), which in Fefferman-Graham coodinates read
 \begin{equation}
t=g(v)\,,\qquad\qquad v=\frac{4-y}{4+y}e^{-ix_-}\,.
\end{equation}
If $t$ is a periodic function, it can be formally expanded in integer powers of $v$, as well as the zero-mode combination $x_+ + x_-$.
The powers of $v$ are easy to identify in the classification above, namely
\begin{equation}
 v^\bh=t^-_{0,\bh,-1}\label{singletonmodes}
\end{equation}
We will restrict our attention to the case of positive powers, $\bar h \geq 0$, which are regular in the `center' of AdS$_3$ which is at $y=4$ in our coordinates.
We would like to study the smallest representation of $\mathrm{SL}(2,\mathbb{R})\times \overline{\mathrm{SL}(2,\mathbb{R})}$ that contains the states (\ref{singletonmodes}). It consists of
all the states which can be obtained by acting with the generators on
\be
t^-_{0,1,-1}= v,
\ee
and we will  denote the resulting vector space by $\calk$.
This carries an infinite dimensional representation, whose structure we will discuss in more detail in a moment, but it is still much smaller than the representation one would obtain by acting on an arbitrary starting point $t^-_{h,\bh}$. This because $L_+$ vanishes on $t^-_{0,\bh,-1}$ while it doesn't vanish on $t^-_{h,\bh}$ when $h\neq 0$, in summary $\calk$ is a short representation.

The representation carried by the vector space $\calk$ is once again reducible but indecomposable:
acting with $L_{-1}$ connects to the $t^+$ modes, i.e. $L_{-1}t^-_{0,\bh,-1}=-\bh t^+_{1,\bh}$, while $L_1 t^+_{1,\bh}=0$.
 \begin{figure}
\begin{center}
\includegraphics[scale=0.6]{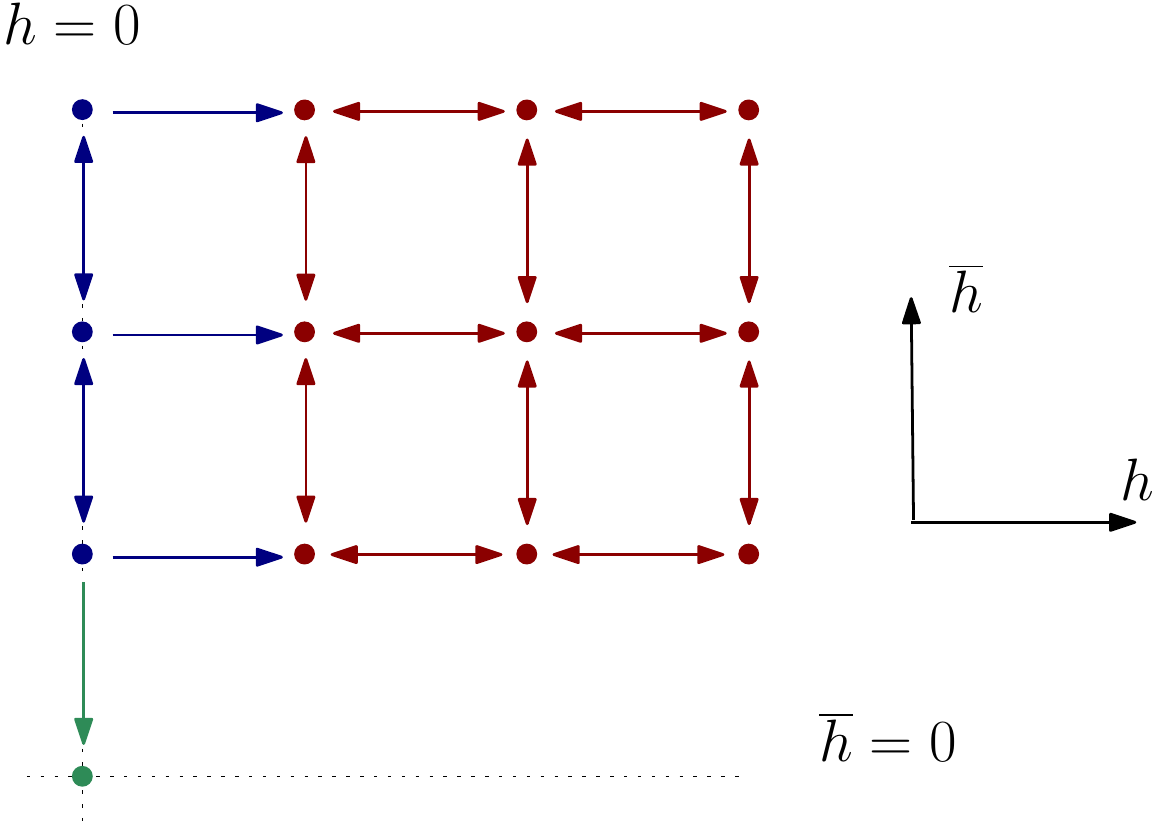}
\end{center}
\caption{A diagram showing the smallest $\mathrm{SL}(2,\mathbb{R})\times \overline{\mathrm{SL}(2,\mathbb{R})}$ representation which contains the solutions (\ref{singletonmodes}). The solutions on the $y$-axis are $t^-$ modes, while the solutions in red are $t^+$ modes.}\label{Kfigure}
\end{figure}
This structure is illustrated in figure \ref{Kfigure} and can be schematically summarized as
\be\begin{array}{ccc}
{\color{blue} (0,1)} & \rightarrow & {\color{red}(1,1)}\\
\downarrow  & &\\
{\color{green} (0,0)} &  & \\
\end{array}.\label{tmintable}\ee
The vector space obtained by acting with symmetry generators on  $t^+_{1,1}$ forms an invariant subspace $\caln$ shown in red,  whose complement in $\calk$ is not invariant.
We remark that by acting by a parity transformation $x_\pm \to x_\mp$ on the above representation we get the smallest representation  containing the solutions
\be
u^h=t^-_{h,0,-1}, \qquad u=\frac{4-y}{4+y}e^{-ix_+,} h \in \NN\label{singletonmodesu}
\ee
The invariant subspace in this representation is again  $\caln$, which is parity invariant.

We note that  the representation on $\calk$,  is non-unitary, since it follows from the algebra that the all states in $\caln$ should have zero-norm in a unitary representation \cite{Flato:1990eu}. The standard way around this to  consider instead the quotient space $\calk/\caln$, on which the induced representation is unitary. In physics terms,  one only considers the $t^-$ part of a solution as physical, while any $t^+$ component of a solution is considered a gauge artifact and hence unphysical. The result of the quotienting procedure is to keep only the singleton representation 
$(0,1)$ spanned by the modes (\ref{singletonmodes}), which we will henceforth refer to as the singleton modes.  The standard physical realization of
this quotienting is to introduce a suitable gauge symmetry. Remarkably, this can  be accomplished by {replacing} the equation of motion (\ref{eomt}) by the fourth-order equation\footnote{Note that here the leading term is fourth order, so it is not simply a quadratic theory plus higher derivative corrections, which is the typical situation.} $\Box^2 t=0$ as discussed\footnote{Since this equation  is parity-invariant, the resulting theory  also contains the $(1,0)$ singleton representation carried by the solutions (\ref{singletonmodesu}).}  in \cite{Flato:1990eu}.  Here, we will not consider such a modification of the bulk Lagrangian but rather investigate which consistent boundary conditions can be imposed on the theory so as to include the singleton modes (\ref{singletonmodes}).

We end with a remark which will be important in what follows: even though the singleton modes $v^{\bar h}$ are part of a nondecomposeable representation from the point of view of the full symmetry group $\mathrm{SL}(2,\mathbb{R})\times \overline{\mathrm{SL}(2,\mathbb{R})}$, they carry  a unitary irreducible representation of the subgroup  U(1)$\times \overline{\mathrm{SL}(2,\mathbb{R})}$, where the U(1) is generated by $L_0$. This representation can be denoted as $(1)_0$, where the subscript refers to the U(1) charge.
We will see below that the singleton modes $ v^\bh$ naturally fit within boundary conditions which break $\mathrm{SL}(2,\mathbb{R})$ down to U(1).
Similarly, the parity-related set of modes  (\ref{singletonmodesu}) also carries  a unitary irreducible representation, but of  a different subgroup, namely $\mathrm{SL}(2,\mathbb{R})\times  \overline{\mathrm{U(1)}} $.

\subsubsection{Zero-mode representations}
So far  we didn't include   the zero-mode solutions (\ref{zeromodes}) in our discussion of representation theory.
Acting with the symmetry generators on them, they connect to both the $t^+$ modes and the singleton modes discussed above. For example, acting with generators on $t = x_-$ we obtain
\begin{align}
L_0 x_- &= 0, & \bar L_0 x_- &= i t^-_{0,0}\\
L_\pm x_- &= -{i\over 2} t^+_{\mp 1,0}, & \bar L_\pm x_- &= i t^-_{0,\mp 1,0}.\label{zeromultipl}
\end{align}

\subsection{Variational principle}\label{parvariation}

So far we have discussed the full solution space of the massless scalar, without imposing any boundary conditions. If we impose boundary conditions which are $\mathrm{SL}(2,\mathbb{R})\times \overline{\mathrm{SL}(2,\mathbb{R})}$ invariant, they will select certain subrepresentations within the full solution space.  In the next sections we will investigate which subspaces of solutions
are selected by imposing consistent boundary conditions, focusing especially on the boundary conditions which allow the singleton modes (\ref{singletonmodes}), and discuss their representation content.
In investigating possible boundary conditions we require the usual physical conditions motivated by holography: the boundary conditions should lead to a {\em consistent variational principle}, meaning that the variation of the action is proportional to the equations of motion without additional boundary terms, and that this {\em action  is on-shell (os) finite}. Our starting point is the bulk action \eqref{SFree} and we will now investigate which possible boundary terms can be added to realize these physical requirements.

As is well known,  the bulk action (\ref{SFree})  is divergent when evaluated on generic solutions due to the infinite volume of AdS$_3$. One regularizes this infrared divergence  by putting the boundary at $y = \e$, for some small $\e$, we'll denote this clipped AdS$_3$ space by $\calm_\e$.
One can then compute that the divergent part of the on-shell action is
\be
S_\mathrm{bulk} \stackrel{\mathrm{os}}{=} \frac{l }{2} \log \epsilon   \Re\int_{\partial\calm_\e}  dx_+dx_-\,  t_0 \partial_+\partial_-\bar t_0 + {\rm \ finite}
\ee
The divergence can be cancelled by adding a boundary term of the form
\begin{equation}\label{Sbnd}
S_{\mathrm{bnd}}=-\frac{l}{2}\Re\int_{\partial\calm_\e}dx_+dx_-\,(\log \epsilon\,t_0\partial_+\partial_-\bar t_0+\call_{\mathrm{bnd\,0}})\,,
\end{equation}
where the boundary Lagrangian density $\call_{\mathrm{bnd\,0}}$ is a function of the boundary fields $t_0, t_2$ and their derivatives that remains finite in the $\epsilon\rightarrow 0$ limit.

The action $S_\mathrm{tot}=S_\mathrm{bulk}+S_\mathrm{bnd}$ is thus by definition the most general boundary extension of \eqref{SFree} that is on-shell finite. We can then investigate the second requirement of consistency of the variational principle. One computes that
\begin{equation}\label{deltaS}
\delta S_\mathrm{tot}\stackrel{\mathrm{os}}{=}\frac{l}{2}\Re\int_{\partial\calm_\e}dx_+dx_-\, \left(2(t_2+\partial_+\partial_-t_0)\delta \bar t_0- \delta \call_{\mathrm{bnd\,0}}\right)+\calo(\epsilon\log\epsilon)
\end{equation}
By definition of a consistent variational principle the above should vanish. This then gives us the following relation between the choice of boundary Lagrangian and choice of boundary conditions on the fields:
\begin{equation}\label{bndrel}
\delta \call_{\mathrm{bnd\,0}}\stackrel{\mathrm{os}}{=}2\Re\left[(t_2+\partial_+\partial_-t_0)\delta \bar t_0\right]\quad
\end{equation}
In what follows  we will discuss  different solutions to the above equation and their  representation content.

\subsection{Example: Dirichlet boundary conditions}
We start by reviewing the standard Dirichlet boundary condition, where $t$ is fixed on the boundary:
\be\label{diri}
\d t_0 = 0
\ee
This provides a solution to (\ref{bndrel}) for the trivial choice $\call_{\mathrm{bnd}\,0}=0$.

Assuming that  $t_0$ is single-valued, we can expand it in Fourier coefficients as
\begin{equation}
t_0=\sum_{h,\bh} c_{h,\bh}e^{-i(hx_++\bh x_-)} + \tilde c_0 ( x_+ + x_-)
\end{equation}
which via our classification above implies that the most general solution satisfying the boundary condition \eqref{diri} is
\begin{equation}
t=\sum_{h,\bh} c_{h,\bh}t_{h,\bh}^-  + \tilde c_0 ( x_+ + x_-)+\sum_{h,\bh} a_{h,\bh} t_{h,\bh}^+,\qquad\qquad a_{h,\bh}\ \mbox{arbitrary\,.} \label{Ddecomp}
\end{equation}

If we want to preserve the full $\mathrm{SL}(2,\mathbb{R})\times \overline{\mathrm{SL}(2,\mathbb{R})}$ symmetry, we must restrict $t_0$ to be invariant under the action of the  symmetries  in (\ref{rep1}-\ref{rep2});
 this is only the case for constant $t_0$, i.e.
\be
t_0= c_{0,0}\,. \label{standardbc}
\ee
With this boundary condition a basis for the set of allowed solutions is given by the $t^+$ modes.  These form the indecomposable representation which was discussed in paragraph \ref{parDmodes}, see figure \ref{tplusfig} and (\ref{tplustable}). If in addition
one imposes regularity in the interior of AdS one ends up with  four unitary\footnote{Meaning that for each of these representations we can find an inner product with respect to which it is unitary, they will not all be unitary with respect to the same inner product however. Note that there are no unitary representations with $L^2=\bar L^2=0$ and non-integer weights \cite{Balasubramanian:1998sn}.} representations built on top of the solutions $t^+_{\pm1,\pm1}$ and $t^+_{\pm 1,\mp1}$.
One can also show that the boundary condition (\ref{standardbc}) is invariant under the extension of the global symmetry algebra to the asymptotic Virasoro symmetry Vir$\times \overline{\rm Vir}$.


There is also a standard and simple holographic interpretation of the boundary condition \eqref{diri}. From the CFT point of view we added an operator
\begin{equation}
\Delta S_\mathrm{CFT}=\frac{l}{2} \int dx_+dx_- t_0 \calo\label{CFTshift}
\end{equation}
and when $t_0$ satisfies the condition \eqref{standardbc} the theory is conformally invariant, so that $\calo$ must be a weight $(1,1)$ primary operator. Therefore, changing the fixed constant value of $t_0$ corresponds to marginally deforming  the dual CFT. Using the standard holographic dictionary we can compute the VEV of the dual operator via the variation of the classical action of the bulk theory:
\be
\langle\calo\rangle={\d S_{\mathrm{tot}} \over \d t_0} =  l   \bar t_2\,.
\ee

Note that if we don't impose conformal invariance and allow  $t_0$ to be a non-constant function, the dual interpretation is that of a state in a theory deformed  by a term   (\ref{CFTshift}) which breaks conformal invariance. In this sense the singleton modes (\ref{singletonmodes}) can be interpreted within Dirichlet boundary conditions, as in the analysis of \cite{Raeymaekers:2015sba}, but each mode represents a state in a different theory.

\section{Conformal boundary conditions for singleton modes?}\label{secnogo}

Rather than interpreting each singleton mode (\ref{singletonmodes}) in a different theory, we would like to find boundary conditions which allow {\em all} the singleton modes, so that they represent different states in
  the same theory. In the mass range (\ref{window}), this would be similar to the Neumann boundary condition imposed in the alternate quantization,  but because the massless scalar is on the boundary of this range a separate analysis needs to be made.  We will argue in this section that there are no consistent SL$(2,\mathbb{R})\times \overline{\mathrm{SL}(2,\mathbb{R})}$ invariant boundary conditions which allow the full class of solutions  (\ref{singletonsols}), unless we take a scaling limit of the scalar theory which keeps only the extreme infrared modes, as was advocated in \cite{Ohl:2012bk}.
On the other hand, we will argue in the next section  that there exist consistent boundary conditions preserving the subgroup U(1)$\times \overline{\mathrm{SL}(2,\mathbb{R})}$ and which do allow the solutions (\ref{singletonsols}).

\subsection{A no-go argument}
In section \ref{parsingmodes} we saw that the singleton modes are part of  a space $\calk$ which carries a  nonunitary and nondecomposable representation as illustrated in figure \ref{Kfigure} and (\ref{tmintable}).
Putting worries about unitarity aside for the moment,
we would like to address the following question: Do there exist conformally invariant boundary conditions, which are consistent in the sense discussed in section \ref{parvariation}, and which  allow the states of $\calk$? 

The problem, as we will now see, lies not in the existence of conformal boundary conditions but in their consistency with a  variational principle. From the near-boundary expansion (\ref{basisas}) we see that all the solutions in
$\calk$ satisfy
\begin{equation}
\partial_-\partial_+t_0=0\label{Kcond}.
\end{equation}
while the more general $t^-_{h,\bar h}$ solutions with $h,\bar h \neq 0$ do not.
This is therefore a natural conformally invariant boundary condition to impose in order to allow for the solutions in $\calk$.  Because (\ref{Kcond}) is parity and charge-conjugation invariant it includes also the representations
related to $\calk$ by these discrete symmetries. It furthermore includes all the Dirichlet $t^+$ modes illustated in figure \ref{tplusfig}: both the red modes (which lie in $\caln$ and its images under discrete symmetries)
and  the modes on the horizontal and vertical axes.

In order for the boundary condition (\ref{Kcond}) to be consistent with a variational principle for an action that is on-shell finite, we have to find a boundary Lagrangian which satisfies \eqref{bndrel}. In this case this  reduces to the requirement\footnote{Note that this requirement is non-trivial as $t_2\neq0$ at least for some solutions if one wants conformal invariance. Indeed, imagine imposing $t_2=0$, this would select the states $t^-_{h,0,0}$ and $t^-_{0,\bh,0}$ only. Rewriting $t^-_{h,0,0}=\frac{1}{2}(t^-_{h,0,1}+t^-_{h,0,-1})$ it follows from \eqref{rep2} that acting with $\bar{L}_{\pm}$ would generate a $t^+$ state which has non-zero $t_2$.}
\begin{equation}
\delta \call_{\mathrm{bnd\,0}}\stackrel{\mathrm{os}}{=}t_2\delta \bar t_0+\bar t_2\delta \bar t_0,\label{nogoeq}
\end{equation}
but one realizes rather directly that there exists no $\call_{\mathrm{bnd\,0}}$ that can satisfy the above constraint.  Indeed, we can formally identify the variation $\delta$ with an exterior derivative on the space of boundary fields. The fact that \eqref{nogoeq} has no solutions is then the simple observation that the LHS is exact and hence closed, while the {RHS} is not closed. In other words applying $\delta$ to $\eqref{nogoeq}$ we find
\begin{equation}
\delta t_2\wedge\delta \bar t_0+\delta\bar t_2\wedge\delta \bar t_0\stackrel{\mathrm{os}}{=}0
\end{equation}
But from our classification of the on-shell solutions we see that even in the presence of the boundary condition \eqref{Kcond}, $t_2$ remains an arbitrary complex  function on-shell. Of course \eqref{Kcond} implies that $\partial_-\partial_+\delta t_0=0$ on-shell, but this still has an infinite number of independent solutions. In other words all of $\delta t_2$, $\delta \bar t_2$, $\delta \bar t_1$, $\delta t_1$ remain linearly independent even on-shell and so \eqref{nogoeq} is false, implying there exists no boundary completion of the bulk action \eqref{SFree} that is on-shell finite and has a variational principle consistent with the boundary condition \eqref{Kcond}.

Of course, while (\ref{Kcond}) seems to be  a  natural boundary condition which selects the states in $\calk$, we have not  proven that there doesn't  exist a different boundary condition which is consistent with a variational principle.
\subsection{Comments on the extreme IR limit}
In making our no-go argument, we have focused on  modes of the field whose energy doesn't scale with the IR cutoff $\e$ as we take $\e$ to zero. The no-go conclusion can be avoided if we instead take a type of extreme IR
limit of the scalar theory, in which only boundary modes survive, and is essentially the  limit considered in \cite{Ohl:2012bk}. This limit is obtained by considering the bulk action (\ref{SFree}) without adding any boundary terms, i.e. we take $S_{\mathrm{bnd}}=0$ instead of (\ref{Sbnd}). Because of the divergent boundary term, the solutions for which the on-shell action is finite are the modes of the rescaled field
\be
w = \sqrt{|\log \e|}\, t
\ee
which stay finite finite as $\e \to 0$. From (\ref{deltaS}) the variation of the action is
\begin{equation}\label{deltaSIR}
\delta S_\mathrm{tot}\stackrel{\mathrm{os}}{=}l\Re\int_{\partial\calm_\e}dx_+dx_-\, \left(\partial_+\partial_-w_0 + {w_2+ \partial_+\partial_-w_0 \over|\log \e|}\right)  \delta \bar w_0 +\calo(\epsilon\log\epsilon)
\end{equation}
This vanishes in the limit $\e\to 0$ if we impose the boundary condition analogous to (\ref{Kcond})
\begin{equation}
\partial_-\partial_+w_0=0.
\end{equation}
which therefore allows for the (rescaled) solutions in $\calk$.
It was shown in \cite{Ohl:2012bk} that the natural norm  of states in $\caln$ goes to zero in the limit, thereby keeping only the singleton representations. Note that in this limit, the rescaled field lives purely on the boundary and doesn't backreact on the bulk metric.
\section{Chiral boundary conditions}\label{secchiral}
In the previous section we  argued that, unless one takes the IR scaling limit discussed above, it is not possible to impose consistent boundary conditions which preserve the
full $\mathrm{SL}(2,\mathbb{R})\times \overline{\mathrm{SL}(2,\mathbb{R})}$ symmetry as well as include the singleton modes of the form $t = v^{\bar h}$ with $\bar h> 0$. This suggests that including these modes in  the two-derivative bulk theory\footnote{As outlined in paragraph \ref{parsingmodes}, a way to obtain a theory of the singleton modes
is to modify the bulk Lagrangian to a fourth-order one. In this theory, which we will not consider here,   both the $(1,0)$ modes $t = u^h$ and the  $(0,1)$ modes $t = v^{\bar h}$ are  physical while the $(1,1)$ modes in (\ref{tmintable}) are pure gauge.}  will   entail breaking at least part of the symmetry. We also observed that the singleton modes  do carry
an irreducible representation $(1)_0$ of the subgroup U(1)$\times \overline{\mathrm{SL}(2,\mathbb{R})}$.

In the first subsection below we will show that there exist consistent boundary conditions,
which arise from a boundary term which breaks $\mathrm{SL}(2,\mathbb{R})$ to U(1), which allow for the singleton modes.  We will then go on to show that these boundary conditions  are a generalization of the `chiral'  boundary conditions for pure gravity studied in \cite{Compere:2013bya} and extend to an asymptotic symmetry algebra which includes both a   right-moving U(1) current algebra and a  right-moving Virasoro algebra. After, this, we extend the discussion to include  gravitational backreaction  and show that our scalar boundary conditions can be consistently combined with the boundary conditions on the metric discussed in \cite{Compere:2013bya} to describe fully backreacted singleton modes.

\subsection{Chiral boundary conditions}
Without further ado, we {propose the following} chiral
 boundary conditions:
\begin{equation}
B_1 \equiv \partial_+ t_0=0,\qquad B_2 \equiv t_2+\frac{i}{2}\partial_- t_0=0\label{CRB}
\end{equation}
It is straightforward to check that they allow general solutions of the type $t = g(v)$. In terms of the mode solutions discussed in section \ref{secfree}, our boundary conditions include:
\begin{itemize}
\item the solutions $t^-_{0,\bh,-1}= v^{\bar h}$ which  for $\bh >0$ are the singleton modes.
\item a linear combination of the zero-mode $x_-$ and the mode $t^+_{0, 0}$, namely
\be
x_- - {i \over 2} t^+_{0,0} = i \ln v.\label{logv}
\ee
Note that this is an allowed field configuration only if the real part of $t$ is compact.
\end{itemize}
It then directly follows from the action of the generators \eqref{rep1} and (\ref{zeromultipl}) that the $L_{-1}$-symmetry is broken by these boundary conditions while the other generators $L_0, L_1, \bar L_m$ are compatible with them, however with $L_1$  acting trivially on all states.

 To obtain a consistent variational principle one can add to the regularized bulk action the boundary term
\begin{equation}
\call_{\mathrm{bnd}\,0}=\half \Im\, \bar t_0\partial_-t_0\label{bdytermAdS}
\end{equation}
The consistency and finiteness follows from observing that this boundary Lagrangian satisfies \eqref{bndrel} when the boundary conditions \eqref{CRB} are imposed.
Indeed, from (\ref{deltaS}), we find for the variation of the total action
\begin{equation}\label{deltaSchir}
\delta S_\mathrm{tot}\stackrel{\mathrm{os}}{=}l\Re\int_{\partial\calm_\e}dx_+dx_-\, \left(t_2+{i \over 2 } \pa_- t_0 +\partial_+\partial_-t_0\right) \delta \bar t_0+\calo(\epsilon\log\epsilon)
\end{equation}
which indeed vanishes when (\ref{CRB}) holds.

\subsection{Example: multiple vortices}\label{secglobalmon}
As we will explore in more detail in section \ref{secparticles}, the boundary conditions (\ref{CRB}) allow for a class of interesting topological defects in the bulk. Indeed, suppose that $t$ has a periodicity
\be
t \sim t + 1,
\ee
{we can then} consider topological line defects located at constant $v$, around which $t$ has a nontrivial monodromy. In terms of the single-valued
field $M = e^{2\p i t}$, these are vortices for the global U(1) charge. More concretely, such solutions look like
\be
t =  -{i\over 2 \p} \sum_i q_i \ln{v-v_i\over 1 - \bar v_i v}\label{tmonsol}
\ee
where $v_i$ are the locations of the defects and   $q_i$ are the corresponding integer winding numbers. For later convenience, we have introduced image charges such that $\Im t =0$ on the boundary $|v|=1$.
Generically these solutions are linear combinations of the constant mode and the singleton modes $v^\bh, \bh > 0$, with coefficients obtained  by expanding (\ref{tmonsol}) around $v=0$. In the particular case that a defect is present in the center $v=0$ of AdS$_3$, the expansion
includes the mode (\ref{logv}) as well.

\subsection{Asymptotic symmetry algebra}
We now address the question whether our boundary conditions (\ref{CRB})  preserve an extended asymptotic symmetry algebra. It will turn out that they are a natural   extension of the chiral
 boundary conditions proposed in \cite{Compere:2013bya}  for pure AdS$_3$ gravity to include matter. These boundary conditions preserve the subgroup U(1)$\times \overline{\mathrm{SL}(2,\mathbb{R})}$ of the global AdS symmetry,
 which gets extended to a combined right-moving Kac-Moody and right-moving
Virasoro algebra.

As a first order check,  we will now verify that our boundary conditions (\ref{CRB}) for the scalar in a fixed AdS background are preserved by the asymptotic symmetries of \cite{Compere:2013bya}. In the next section we will extend the analysis to include dynamical gravity. The asymptotic symmetries of \cite{Compere:2013bya} are generated by two sets of asymptotic Killing vectors, depending on two arbitrary rightmoving functions $\bar U (x_-), \bar V(x_-)$:
\bea
\h_{\bar U } &=&\bar U \pa_+ +\ldots \label{asU1}\\
\x_{\bar V} &=&\bar V' y \pa_y + \bar V \pa_- + \half \bar V '' y \pa_+ + \ldots\label{asvir}
\eea
 The $\h_{\bar U}$ vectors generate a  right-moving $u(1)$ current algebra, while the $\x_{\bar V}$  are the standard right-moving Virasoro generators.
The near-boundary components of the scalar transform as follows:
   \begin{align}
 \d_{\bar U} t_{0} &= \bar U  \pa_+ t_{0},&  \d_{\bar V} t_{0} &= \bar V \pa_- t_{0} \\
  \d_{\bar U} t_{2} &= \bar U  \pa_+ t_{2},&   \d_{\bar V} t_{2} &= \pa_- ( \bar V t_{2}) + \bar V' \tilde t_{2} + \half \bar V'' \pa_+ t_{0} \\
\d_{\bar U}\tilde  t_{2} &= \bar U  \pa_+\tilde t_{2},&     \d_{\bar V} \tilde t_{2} &= \pa_- ( \bar V t_{2}) .
   \end{align}
Using these expressions, one easily checks that the boundary conditions (\ref{CRB}) are invariant under the asymptotic symmetries generated by  (\ref{asU1},\ref{asvir}), i.e.
\be
\left. \d_{\bar U} B_i \right|_{B_{1,2}=0} = 0, \qquad \left. \d_{\bar V} B_i \right|_{B_{1,2}=0} = 0.
\ee

Before we go on to discuss the inclusion of dynamical gravity, we end with a remark. The attentive reader may have noted  from (\ref{deltaSchir}) that, from    the point of view of the variational principle, we could have
just as well imposed the weaker boundary condition
\be
\tilde B \equiv t_2+{i \over 2 } \pa_- t_0 +\partial_+\partial_-t_0 =0\label{weakerbc}
\ee
which would also have allowed the singleton modes.
The reason we adopted the stronger conditions (\ref{CRB}) is that (\ref{weakerbc}) is not invariant under the symmetries generated by (\ref{asU1},\ref{asvir}), since
\be
\left. \d_{\bar U} \tilde B \right|_{ \tilde B =0}  = {i\over 2} \bar U' \pa_+ t_{0}, \qquad
\left. \d_{\bar V}  \tilde B  \right|_{ \tilde B =0} = \half \bar V'' \pa_+  t_{0} + \bar V' \pa_+ \pa_- t_0 .
\ee

\subsection{Including gravity}
We will now show that the boundary conditions on the scalar (\ref{CRB}) can be extended to the case where backreaction of the scalar on the metric is included, in a consistent manner
and preserving the asymptotic symmetries generated by (\ref{asU1},\ref{asvir}). The boundary conditions on the metric will be those proposed for the pure gravity case in \cite{Compere:2013bya}. The fact that these remain consistent upon inclusion of the massless scalar is  nontrivial, since the scalar backreaction turns on a logarithmic mode in the metric which is absent in the  case of pure gravity. We will here only sketch the derivation, referring to Appendix \ref{appholren}  for more details {and a generalization to a family of axion-dilaton like kinetic terms and} the inclusion of a U(1) Chern-Simons field. {Both these generalizations} will be {relevant} in the next section.

The starting point is the Einstein-Hilbert action for gravity minimally coupled to a massless complex scalar:
\be
S =  \int_\calm \left[  d^3 x \sqrt{- G} \left( \calr + {2 \over l^2} - \half {\pa_\m t \pa^\m \bar t} \right)
 - 2 \int_{\pa \calm} \sqrt{-\g} K \right].\label{Stgrav}
\ee
We have included the standard Gibbons-Hawking boundary term to compensate for the fact that the Einstein-Hilbert term contains second derivatives of the metric.
We use Fefferman-Graham coordinates in terms of which the metric {reads}
\be
ds^2_3 = l^2 \left( {dy^2 \over 4 y^2} + {1 \over y} g_{ij} (x^k,y) dx^i dx^j\right)\label{FGcoords}
\ee
The near-boundary expansion of the fields is then \cite{de Haro:2000xn}
\bea
t &=& t_{0} + y\,t_{2} + y\log y\, \tilde t_{2}+ \calo ( y^2 \log y)\\
g_{ij} &=& g_{0\, ij}+ y\, g_{2\, ij}+ y\log y\, \tilde g_{2\, ij} + \calo ( y^2 \log y).\label{FGmetric}
\eea
As was the case for the scalar field, the logarithmic term in the expansion of the metric is special for the massless scalar and would be absent for a scalar with mass in the window
(\ref{window}).
Expanding the  equations of motion following  from (\ref{Stgrav}) near the boundary, one finds that   the boundary values $t_{0}$ and $ g_{0\, ij}$ are arbitrary, while the logarithmic coefficients $\tilde t_{2}, \tilde g_{2\, ij} $ as well as the trace and the divergence of $g_{2\, ij}$ are fixed in terms of
them:
\begin{align}
\tilde t_{2} &= -{1\over 4} \Box_{0} t_{0}, &
\tilde g_{2\,ij} &= - {1  \over 4} {\pa_{(i} t_{0} \pa_{j)} \bar t_{0}}
+{1\over 8} {\pa_k t_{0} \pa^k \bar t_{0}}g_{0\,ij} \label{eomnbmetr}\\
 g_{2} &= - \half R_{0} +  {1\over 4} {\pa_i t_{0} \pa^i \bar t_{0} }, &
 \nabla_0^j ( g_{2\, ij}- g_{2} g_{0\, ij}) &=\Re\left(  { \bar t_{2} \pa_i t_{0}  } \right).\label{g2sol}
 \end{align}
 Here and in what follows, indices are raised and contracted with the $g_0$ metric.
Once again we must  add boundary terms to the action in order for it to be finite as the IR cutoff $\e$ is taken to zero. This leads to the following generalization of (\ref{Sbnd}):
\be
  S_{\mathrm{bnd}} = {l}\int_{\pa \calm_\e}  d^2x\sqrt{- g}\left[ - {2 \over \e }+  \left( {R \over 2} - {\pa_i t \pa^i \bar t\over 4 } \right)\log \e - \call_{\mathrm{bnd\,0}} \right]+\calo(\epsilon\log\epsilon) \\
  \ee
  where $\call_{\mathrm{bnd\,0}}$ is a finite part which may  depend on both  the metric and the scalar field.
  For the variation of the action one finds
\begin{equation}\label{deltaSmetr}
\delta S_\mathrm{tot}\stackrel{\mathrm{os}}{=}{l }\int_{\pa \calm_\e}  d^2x\sqrt{- g_{0}}\left[2\Re \left( (t_{2}+\tilde t_{2})\d \bar t_{0}\right) -  \left( g_{2\,ij}+ \tilde g_{2\,ij} -g_{2}g_{0\, ij} \right)\d  g_{0}^{ij} -\d  \call_{\mathrm{bnd\,0}}\right]
 +\calo(\epsilon\log\epsilon)
\end{equation}

We now present our proposed boundary conditions for the scalar and the metric, which allow for the backreacted singleton solutions. These combine the scalar boundary condition (\ref{CRB})  with the boundary conditions of \cite{Compere:2013bya} on the metric:
\begin{align}
t_{0} &= t_{0} (x_-), & t_{2} &=-{i\over 2} t_{0}'\label{CRBmetr1}\\
 g_{0\,--} &= \bar P'(x_-), & g_{2\,++} &= {\D \over k}\label{CRBmetr2} \\
  g_{0\,+-} &= -\half, & g_{0\,++} &= 0.\label{CRBmetr3}
\end{align}
Here, $\bar P (x_-)$ is an arbitrary right-moving function, while $\D$ is a fixed  number; different values of $\D$ correspond to different theories.  We also defined
 \be
 k = 4\p \ell = {c\over 6}
 \ee
Note the similarity  between the first two lines  (\ref{CRBmetr1}) and (\ref{CRBmetr2}): as for the scalar, one of the leading components of the metric, $g_{0\,--}$,  is allowed to be a fluctuating right-moving function, while one of the subleading components, $g_{2\,++}$, is fixed.
 Furthermore, the equations  following  from (\ref{Stgrav})  imply
 \begin{align}
 \tilde t_{2} &=0, &
 \tilde g_{2\ij} &= - {1\over 4} |t_{0}'|^2\d_i^- \d_j^- \\
  g_{2+-} &= -{\D \over k} \bar P',&
  \pa_+   g_{2\,--} &= 0
 \end{align}
 As in \cite{Compere:2013bya}, we will parametrize $g_{2\,--}$ in terms of another arbitrary right-moving function
 $\bar L(x_-)$ as
 \be
 g_{2\,--} = {1\over k} \left(\bar L(x_-) + \D (\bar P')^2\right).\label{defL}
 \ee

The boundary conditions (\ref{CRBmetr1}-\ref{CRBmetr3}) are consistent: indeed,  taking the  boundary Lagrangian in (\ref{deltaSmetr}) to be
\be
 \call_{\mathrm{bnd\,0}} = - {\D \over k} g_0^{++} +\half \Im\, ( \bar t_0\partial_-t_0 )
 \ee
 we find that the variation (\ref{deltaSmetr}) vanishes when the boundary conditions (\ref{CRBmetr1}-\ref{CRBmetr3}) hold.

One can check that, as promised, the boundary conditions (\ref{CRBmetr1}-\ref{CRBmetr3}) are invariant under the asymptotic symmetries generated by (\ref{asU1},\ref{asvir}), with the free right-moving  functions transforming as
 \begin{align}
    \d_{\bar V} \bar P'  &= \pa_- ( \bar V\bar P')  & \d_{\bar U} \bar P'  &= - \bar U'  \\
     \d_{\bar V} \bar L  &= \bar V\bar L' + 2 \bar V' \bar L - {k\over 2} \bar V''' - {k\over 4} \bar V' {|t_{0}'|^2 }& \d_{\bar U}\bar L  &= 0\\
      \d_{\bar V} t_{0} &= \bar V  t_{0}'  & \d_{\bar U} t_{0} &= 0
   \end{align}

\section{Examples with point-like sources}\label{secparticles}
As we already noted in paragraph \ref{secglobalmon}, the chiral boundary conditions (\ref{CRB}) on the complex scalar allow for multi-{vortex} solutions  which would not fit in the standard conformally invariant boundary conditions. In this section we explore the holographic meaning of what is essentially the backreacted version of the  solutions of paragraph \ref{secglobalmon}, with the difference that we will now replace the free complex scalar with an interacting axion-dilaton field. Such solutions play an important role in the black hole deconstruction proposal \cite{Denef:2007yt} where they describe wrapped M2-branes, which were proposed as microstate geometries  which make up the entropy of the M-theory black hole of \cite{Maldacena:1997de}. They can also be seen as simple examples of so-called W-branes, the M-theory lift of open strings connecting D6-brane centers, which were conjectured to capture the entropy of a class of D-brane systems \cite{Martinec:2014gka},\cite{Martinec:2015pfa}.

As we shall argue below, in a first approximation these solutions reduce to the geometry produced by multiple conical defects moving on helical geodesics in AdS$_3$, which were considered in \cite{Hulik:2016ifr}.
As a warm-up example, we will therefore reinterpret these solutions in the chiral boundary conditions \cite{Compere:2013bya} for pure gravity, i.e. (\ref{CRBmetr1}-\ref{CRBmetr2}). Interestingly, we will see that these boundary conditions allow a wider class of multi-particle configurations where the total mass exceeds the black hole threshold, already suggesting their relevance for black hole physics.

\subsection{Helical particles}
As explained above, it will be useful to first consider the warm-up example of pure gravity with point-particle sources. We will focus on multi-particle solutions which represent the backreaction of particles moving on helical geodesics in global AdS, which were studied in \cite{Hulik:2016ifr}. These fit in standard Brown-Henneaux boundary conditions when the total mass is below the black hole threshold, but  interestingly we will see that they fit in the more general boundary conditions (\ref{CRBmetr2}-\ref{CRBmetr3}) even when the total mass exceeds the black hole threshold.

The point particle source terms are
\be S_{\rm source} = -  \sum_i {m_i } \int_{W_i} ds_i\label{sourcegrav}
\ee
where $W_i$ are the timelike worldlines of the particle sources and  $m_i$ are the point-particle masses.
Our ansatz for the metric is
\be
ds^2 = l^2 \left[-  (dT + A)^2 + e^{-2\F} dv d\bar v\right]\label{metrintr}
\ee
where the function $\F$ and the one-form $A$ are defined on the base space parametrized by $v,\bar v$.

For example, for global AdS$_3$ these are given by
\be
\F_{\rm AdS} = \ln (1- |v|^2), \qquad A = \Im \left( \pa_v \F_{\rm AdS} dv \right).\label{PhiAdS}
\ee
The relation with standard global coordinates (\ref{globalcoord}) is
\be
v = \tanh \r e^{- i x_-},
\ee
in other words,  $v$  coincides with the holomorphic coordinate we  introduced before in (\ref{singletonsols}).

More generally, the Einstein equations imply that $\F$ must satisfy the Liouville equation\footnote{We will provide more details on the derivation of the equations of motion in the next section.} away from the sources.
 We will take the particles to follow geodesics  of constant $v = v_i$, which corresponds to helical curves $\r =$ constant, $x_- =$ constant in global AdS.
 We then need to solve Liouville's equation in the presence of delta-function\footnote{Our delta function is normalized as $\int d^2 v  \d^2(v, \bar v) = 2 \int d(\Re v)  d(\Im v)   \d^2(v, \bar v)=1$. In particular $\pa_v\pa_{\bar v} \ln |v| = \p \d^2(v, \bar v)$.} sources coming from (\ref{sourcegrav}):
\be
\pa_v \pa_{\bar v}\F + e^{-2\F } = {1\over 4} \sum_i m_i \d^2(v- v_i, \bar v- \bar v_i).\label{Liouv}
\ee
The coordinate $v$ can be taken to run over the unit disk, $|v|\leq 1$, and  to ensure asymptotically AdS behaviour the field $\F$ should satisfy the Zamalodchikov-Zamolodchikov boundary conditions of  \cite{Zamolodchikov:2001ah}
\be
\F = \ln (1- |v|^2) + \calo\left(  (1- |v|^2)^2 \right).\label{zzbc}
\ee
The solution for the one-form $A$ can be expressed in terms of $\F$ as
\be
A = \Im \left( \pa_v \F dv  + d \l  \right)\label{Asol}
\ee
where
\be
\l (v) = - {1 \over 8 \p} \sum_i m_i \ln {v -v_i \over 1 - \bar v_i v}
\ee
The large gauge transformation involving the multivalued function $\l (v)$ ensures that $A$ is free of Dirac string singularities, as required by the equations of motion.

We now show how  solutions satisfying the above constraints fit in the boundary conditions (\ref{CRBmetr2}-\ref{CRBmetr3}). We define the antiholomorphic  Liouville stress tensor
\be
\bar T (\bar v ) = -(\pa_{\bar v}\F)^2- \pa_{\bar v}^2\F.
\ee
Using a doubling trick argument, one can show that $\bar T$ must be of the form
\be
\bar T (\bar v ) =  \sum_{i=1}^N  \left({\e_i \over  (\bar v-\bar v_i)^2}+ {  \e_i \over  (\bar v-1/v_i)^2}+ {\bar c_i\over \bar v- v_i}+ {\tilde {\bar c}_i\over \bar v-1/ v_i}\right)\label{Tintr}
\ee
where the $\epsilon_i$ are related to the particle masses through $\e_i = {m_i\over 8\p} (1- {m_i\over 8\p})$, and the $\bar c_i, \tilde {\bar c}_i$ are called accessory parameters.
Their determination requires solving a certain monodromy problem, and amounts to the knowledge of a particular large $c$ Virasoro vacuum conformal block \cite{Hulik:2016ifr}.   One can show  that the near-boundary behaviour of $\F$ is determined by $\bar T$ as follows:
\be
\F =\ln (1- |v|^2)-{1\over 6} e^{2 i \arg \bar v} \bar T ( e^{ i \arg \bar v}) (1- |v|^2)^2   + \calo \left(  (1- |v|^2)^4 \right).
\ee
Using this and (\ref{metrintr},\ref{Asol}), one finds that a coordinate transformation which brings the metric into the asymptotic form  (\ref{FGmetric}) is
\be
v = \left( 1- {y\over 2} + {y^2 \over 8} \right) e^{-i x_-} + \calo (y^3 ),\qquad
T = \half ( x_+ + x_-) + \calo (y^2).
\ee
The metric can be seen to fit in the boundary conditions (\ref{CRBmetr2}-\ref{CRBmetr3}) with
\be
\D = -{k \over 4}
\ee
where the rightmoving functions $\bar P' (x_-)$ and $\bar L (x_-) $ are given by
\be
\bar P' = - 2 i \pa_- \bar \l(e^{i x_-}), \qquad
\bar L = - {k\over 4} + k e^{2 i x_-} \bar T (e^{i x_-}).\label{chargesdefs}
\ee

Note that, as explained in \cite{Hulik:2016ifr}, a further large coordinate transformation can bring these metrics into Brown-Henneaux boundary conditions, where $g_{0}= - dx_+ dx_-$, provided
that the sum of the point-particle masses is below the upper bound
$ \sum_i m_i \leq 4 \p$. This bound is essentially the black hole threshold: for a single particle in the origin of AdS$_3$, the black hole threshold is $m = 4\p$.
In the current boundary conditions we don't encounter such an upper bound on the  point-particle mass, but the causal structure of the boundary metric $g_0$ naively becomes  problematic when the total point-particle mass exceeds
$ 4 \p$.
It's not clear however how serious a problem this is in the present context:  for standard Brown-Henneaux boundary conditions, the conformal class of the boundary metric, and therefore its causal structure, has physical meaning, but it is not clear what replaces this in the case of chiral boundary conditions.

It will be useful below to work out the explicit solution to first order in an expansion for small masses $m_i$. As shown in Appendix \ref{apppert},  the result is
\be
\F = \ln (1- |v|^2) + {1 \over 4 \p} \sum_i m_i \left( 1 - { |v-v_i|^2 + |1 - \bar v_i v|^2\over |v-v_i|^2 - |1 - \bar v_i v|^2}\ln {|v-v_i|\over  |1 - \bar v_i v|} \right) +\calo( m_i^2 ).\label{soldefs}
\ee
The Liouville stress tensor is, to this order,
\be
\bar T = {1\over 8\p}\sum_i {m_i (1- |v_i|^2)^2 \over (\bar v- \bar v_i)^2 (1 -v_i \bar v)^2}+\calo( m_i^2 ) .
\ee
and  $\bar P'$ and $\bar L $ are given by
\be
\bar P' = -{1\over 4\p} \sum_i m_i {1- |v_i|^2 \over |1- v_i e^{i x_-}|^2  }, \qquad
\bar L = -{k\over 4} + {k \over 8 \p} \sum_i m_i {(1- |v_i|^2)^2 \over |1- v_i e^{i x_-}|^4  }+\calo( m_i^2 )
\ee

The U(1) current algebra and Virasoro charges are defined as in \cite{Compere:2013bya}
\be
\bar \calp_m = {1\over 2 \p} \int_0^{2\p} d\varphi e^{i m x_-} \left( \D + 2  \D  \bar P'\right),\qquad
\bar \call_m = {1\over 2 \p} \int_0^{2\p} d\varphi e^{i m x_-} \left( \bar L - \D (\bar P')^2\right).\label{chiralcharges}
\ee
They  were shown to satisfy the Dirac bracket algebra
\bea
i \{ \bar \call_m, \bar \call_n \} &=& (m-n) \bar \call_{m+n} + {k\over 2}  m^3 \d_{m,-n}\\
i \{ \bar \call_m, \bar \calp_n \} &=& -n \bar \calp_{m+n} \\
i \{ \bar \calp_m, \bar \calp_n \} &=&- 2\D  m \d_{m,-n}.
\eea
Evaluated on our solutions, one finds
\bea
\bar \calp_m &=&  {k\over 4} \left( -\d_{m,0} + {1\over 2 \p} \sum_i m_i (1-|v_i|^2) \sum_{n \geq {\rm sup} \{0,m\} } \bar v_i^n v_i^{n-m}\right)\\
\bar \call_m &=&  {k\over 4} \left(- \d_{m,0} + {1\over 2 \p} \sum_i m_i (1-|v_i|^2)^2 \sum_{n \geq {\rm sup} \{0,m\} } n(n-m) \bar v_i^{n-1} v_i^{n-m-1}\right)+\calo( m_i^2 )\nonu\label{charges1}
\eea
In particular, the zero modes are
\be
\bar \calp_0 =  - {k\over 4} +{k\over 8\p} \sum_i m_i, \qquad \bar \call_0 =  - {k\over 4} +{k\over 8\p} \sum_i m_i +\calo( m_i^2 ).\label{charges2}
\ee

\subsection{W-branes}\label{secW}
The next example we will consider is motivated by black hole physics, more specifically the black hole deconstruction proposal for the construction of microstate geometries in supergravity with M2-brane sources \cite{Denef:2007yt}.
After dimensional reduction to 3 dimensions, these involve AdS gravity coupled to an  axion-dilaton field $\t$ \cite{Levi:2009az},\cite{Raeymaekers:2014bqa},\cite{Raeymaekers:2015sba} rather than the  free complex scalar considered in the previous sections. The bulk action is now
\be
S_{\rm bulk} =  \int_\calm   d^3 x \sqrt{- G} \left( \calr + {2 \over l^2} - {\pa_\m \t \pa^\m \bar \t\over 2  (\Im \t)^2} \right)\label{sbulktau}
\ee
The  axion-dilaton equation of motion following from this action is
\be
 \Box \t + i {\pa_\m \t \pa^\m  \t\over \Im \t}=0.\label{eomtau}
\ee
A necessary condition to trust the effective 3D description (\ref{sbulktau}) is that, near the boundary of AdS$_3$,
 the  $\t$ profile describes a small fluctuation around  a large, constant, imaginary  value:
 \be
 \t = i V_\infty + t,\qquad {|t| }\ll V_\infty, \label{tauVinfty}
 \ee
where $V_\infty$ is the volume of the internal Calabi-Yau space in 11D Planck units.
From (\ref{sbulktau}) and  (\ref{eomtau}) we see that $t$ behaves like a free scalar near the boundary, so that our earlier free scalar  analysis  is a good approximation near the boundary.
Also, one sees that the stringy $SL(2,\ZZ)$ symmetry of the $\t$ field reduces to the periodicity $t\sim t+ 1$ of the small fluctuation.

Furthermore, independent of whether we are near the boundary where (\ref{tauVinfty}) holds,  the equation (\ref{eomtau})
shares, in {a} global AdS background, a set of solutions  with the free scalar field, namely those for which the two terms in (\ref{eomtau}) vanish separately.
This is precisely the case for the singleton solutions
\be
\t = g( v ).\label{singletontau}
\ee
When the bosonic system is embedded in a theory with `left-moving' supersymmetry (i.e. where the left-moving SL(2,$\RR$) factor gets enhanced to a supergroup), as is the case in the black hole deconstruction context, these solutions preserve  the left-moving supersymmetries.

Once again, one can analyze which
 consistent boundary conditions for  axion-dilaton-gravity allow for the solutions (\ref{singletontau}). This proceeds  largely parallel to the free scalar-plus-gravity system and we refer to Appendix \ref{appholren} for details.
The asymptotic behaviour of the fields is unmodified and   the resulting chiral boundary conditions  are again (\ref{CRBmetr1}-\ref{CRBmetr3}), with $t$ replaced by $\t$.

We are interested in solutions in the presence of W-brane source terms, which arise from M2-branes wrapped on an internal two-sphere.  These are point-like objects in   the effective three-dimensional  description which couple both to gravity and the axion-dilaton field in the following way:
\bea
S_{\rm source} &=& - \sum_i {q_i } \int_{W_i} \left[ {ds_i \over \Im \t} - \s  C\right] \label{Wbract}
\eea
where $\s = \pm 1$ is a sign factor, which in our conventions is one for  M2-brane sources and minus one for  anti-M2-branes. The one-form $C$ is dual to the axion which is the  real part of $\t$:
\be
d C = {1\over (\Im \t)^2} \star d (\Re \t )
\ee
and the second term in (\ref{Wbract}) represents a magnetic charge for the axion field.

The equations of motion are
\bea\label{eomsW1}
\calg^{\m\n}-{1\over \ell^2} G^{\m\n}-\half \calt^{\m\n}&=&
 \sum_i {q_i\over 2} \int dw_i { \d^3(x- x_i(w_i)) \over \sqrt{-G} \Im \t } {\dot x_i^\m \dot x_i^\n \over \sqrt{-G_{\r\s} \dot x_i^\r \dot x_i^\s}}\nonu
 d d (\Re \t ) &=& \s \sum_i q_i \int d w_i { \d^3(x-x_i(w_i))\over \sqrt{-G}}\dot x_{i\m} \star d x^m\\
 \Box (\Im \t) + {(\pa(\Re \t))^2 - (\pa(\Im \t))^2 \over \Im \t} &=& - \sum_i  q_i \int dw_i  { \d^3(x-x_i(w_i))\over \sqrt{-G}}\sqrt{-G_{\m\n} \dot x_i^\m \dot x_i^\n}.\nonumber
\eea
where $\calg_{\m\n}$ is the Einstein tensor and $\calt_{\m\n}$ the matter stress tensor
\be
\calt_{\m\n} = {1\over  (\Im \t)^2} \left(\pa_{(\m}\t \pa_{\n)}\bar \t-\half \pa_{\r}\t \pa^\r \bar \t G_{\m\n}\right).
\ee

Our ansatz for the metric is once again (\ref{metrintr})\footnote{Our current conventions are related to those of \cite{Raeymaekers:2015sba} as: $t = t_{there}/2,v = z_{there}, A = \chi_{there}/2, \F = \F_{there}- \ln (\Im \t ) +\ln 2, q_i =2 \p q_{i\ there} $.}, with
$v$ running over the unit disk, and $\F$ should satisfy the Zamolodchikov-Zamolodchikov boundary conditions (\ref{zzbc}).

We will once again take the W-branes to follow geodesics  of constant $v = v_i$,  corresponding to helical curves in global AdS. The equations of motion reduce to
\bea
\pa_v\pa_{\bar v} \F + e^{-2 \F} -{\pa_v \t \pa_{\bar v} \bar \t+\pa_{\bar v} \t \pa_{v} \bar \t \over 8 (\Im \t)^2} &=& \sum_i {q_i \over 4 \Im \t} \d^2(v-v_i,\bar v- \bar v_i)\label{eom3}\\
dA &=& - i e^{-2 \F} dv \wedge d\bar v\label{eom1}\\
\pa_v \t \pa_v \bar \t &=&0\label{eom2}\\
\half ( \pa_v \pa_{\bar v} + \pa_{\bar v} \pa_v )(\t- \bar \t) &=& - i\sum_i q_i \d^2(v-v_i,\bar v- \bar v_i)\label{eom4}\\
\half ( \pa_v \pa_{\bar v} - \pa_{\bar v} \pa_v )(\t+ \bar \t) &=&i \s \sum_i q_i \d^2(v-v_i,\bar v- \bar v_i)\label{eom5}
\eea
The solution for the equation (\ref{eom1}) for $A$  is, up to an exact form which can be absorbed in a redefiniton of $t$, given in terms of $\F$ and $\t$ as
\be
A = \Im \left( \pa_v(\F + \half \ln (\Im \t) ) dv\right) .
\ee
as one can check using (\ref{eom3}-\ref{eom5}).
For $\s=1$, i.e.  for M2-brane sources, the solution of equations (\ref{eom2}-\ref{eom5}) for $\t$ satisfying the  condition (\ref{tauVinfty})  is precisely the multi-centered global {vortex} solution we encountered in paragraph (\ref{secglobalmon}):
\be
\t = i V_\infty -{i\over 2 \p} \sum_i q_i \ln{v-v_i\over 1 - \bar v_i v}.\label{tausol}
\ee
For anti-M2-branes ($\s=-1$), one would find a similar antiholomorphic solution.

Now, let's consider the backreacted metric. Substituting (\ref{tausol}) into (\ref{eom3}), we see that, as long as
  \be
  m_i \equiv {q_i \over V_\infty}\ll 1,
  \ee
  as is the case in the regime of interest, the first order deviation from the AdS background is the same as that produced by point
 particles on helical geodesics with masses $m_i$, discussed in the previous section. The solution for the metric to this order is simply given by
 (\ref{soldefs}), and the conserved asymptotic charges are given by  (\ref{charges1},\ref{charges2}).

\section{Discussion: a refined deconstruction proposal}

In this work we studied  boundary conditions in two-derivative scalar-gravity theories which allow for the class of solutions (\ref{singletonsols}), and the symmetries preserved by them. We introduced chiral boundary conditions
 (\ref{CRBmetr1}-\ref{CRBmetr3}) which include those solutions and follow consistently from a variational principle with a finite on-shell action.  We  also showed that these preserve
the asymptotic symmetries of \cite{Compere:2013bya}, which form a combined left-moving Virasoro and U(1) current algebra.
We didn't however work out the contribution of the scalar field to the asymptotic charges, which for the pure gravity case \cite{Compere:2013bya} were computed in the formalism of \cite{Barnich:2001jy},\cite{Barnich:2007bf}. This contribution starts at the second order in an expansion in the scalar profile, which goes beyond the approximation considered in this work.
Obtaining well-defined charges to all orders may involve a  generalization of our boundary conditions since  the scalar field sources the logarithmic mode in the metric.
The fact that we obtained a finite on-shell action stems us hopeful  that this should be possible. 

Since the analysis in this work was motivated by
the holographic interpretation of
scalar-gravity  solutions which arise in the black hole deconstruction (BHD) proposal, let us summarize what our results imply, in our view, for this proposal. In \cite{Denef:2007yt}, certain D2-brane configurations were proposed to semiclassically represent microstates of a 4D stringy black hole, based on an appealing counting argument in the probe approximation which relies on  their huge lowest Landau level degeneracy.

Upon taking an  M-theory decoupling limit, the backreacted BHD solutions  become essentially the M2-brane solutions of section \ref{secW}, with an additional  constraint that the total  M2-charge should vanish imposed by tadpole cancellation. In order to satisfy this constraint while preserving supersymmetry  one is led to consider solutions of the type (\ref{tausol}) where some of the $q_i$  are negative, so that $\sum_i q_i=0$. Such `negative branes', which have negative tension, were investigated in \cite{Dijkgraaf:2016lym}. In addition, in the BHD configurations  an additional U(1) Chern-Simons field in the 3D theory is switched. This field  doesn't interact with the metric and axion-dilaton fields in the bulk and therefore our solutions for these fields are unmodified, but its presence does influence the boundary theory as is familiar from discussions of holographic spectral flow \cite{David:1999zb},\cite{Maldacena:2000dr}.
In Appendix \ref{appholren}, we show that adding a U(1) Chern-Simons field $\cala$ and imposing suitable boundary conditions on it changes the value of $\D$ in (\ref{CRBmetr2}) as follows:
\be {\D \over k} = g_{2\,++} + {1 \over 4} \cala_{0+}^2  \ee
The  BHD  solutions have \cite{deBoer:2008fk} $ \cala_{0+}=1$ and $ g_{2\,++} = -1/ 4$, and therefore belong to the theory with
\be
\D = 0.
\ee

The same stringy  black hole of the BHD proposal  was of course originally   studied in \cite{Maldacena:1997de}, and its  microstates were identified as states in a dual CFT, the so-called MSW theory. Our work points to some tension in trying to interpret the BHD solutions as the bulk duals of the MSW microstates\footnote{Another objection to such an interpretation was raised in \cite{Tyukov:2016cbz}.}. 
We argued in \ref{secnogo} that,
 unless one takes an IR scaling limit of the theory in which the scalar has no bulk dynamics, it is not possible to fit the BHD solutions  within conformally invariant boundary conditions.
 Therefore it seems hard to maintain that the BHD  solutions are the bulk duals of the black hole microstates in  the MSW CFT or in  a  marginal deformation thereof.

 On the other hand, we observed that the BHD solutions do naturally fit in a generalization of the chiral boundary  conditions of \cite{Compere:2013bya}, which still allow for a right-moving Virasoro algebra together with a U(1) current algebra.  On the bulk side, the new boundary conditions arose from adding additional finite boundary terms to the action,
 and one expects therefore that the dual field theory also arises from some  deformation of the  MSW theory, and it would be interesting to make this more precise.
  Letting components of the boundary metric fluctuate typically means coupling the dual theory to gravity \cite{Compere:2008us,Apolo:2014tua}, but it is at present unclear how  the chiral boundary conditions of
   \cite{Compere:2013bya} arise  in this manner.
  Since the chiral theory still contains an infinite number of conserved charges, it is also possible that it could interpreted as  a type  of `integrable' deformation  \cite{Smirnov:2016lqw} of the MSW theory.

 The proposed chiral boundary conditions still allow for the  extremal BTZ black hole with $M + J=0$, of  which the BHD solutions are the proposed microstates: it is easy to see \cite{Compere:2013bya}  this solution also lives in the
 $\D=0$ theory and has $\bar P'=0, \bar L =M$. These considerations suggest that chiral theory, which includes both the BHD solutions and the extremal BTZ, is the
 proper setting to interpret the BHD proposal holographically.
The main open  question which remains is whether the BHD solutions, which have extra structure or hair encoded into the other Virasoro and current algebra modes, can realize the same zero mode charges as the extremal BTZ geometry, and if so,
if there is  a sufficiently large moduli space \footnote{It will be important to take into account the lowest Landau level degeneracy in the internal space, which results in a large number of M2-particle species in the effective 3D description.} of them to account for the entropy. To answer this question it would also be of interest to study the extra asymptotic structure emerging when our chiral boundary conditions are generalized to the additional fields present in the 3D $(4,0)$   supergravity theory which governs the black hole deconstruction setup.

\section*{Acknowledgements}
It is a pleasure to thank I. Bena, N.S. Deger, M. Guica, O. Hul\'{i}k, M. Porrati,  T. Proch\'{a}zka,  D. Turton and B. Vercnocke for discussions and useful suggestions. The research of JR was supported by the Grant Agency of the Czech Republic under the grant 17-22899S, and by ESIF and MEYS (Project CoGraDS - CZ.02.1.01/0.0/0.0/15\_003/0000437).
 DVdB was partially supported by the Bo\u{g}azi\c{c}i University Research Fund under grant number 17B03P1. This collaboration was supported by the bilateral collaboration grant T\"UBITAK 14/003 \& 114F218.
JR would like to thank the Galileo Galilei Institute for Theoretical Physics (GGI) for the hospitality and INFN for partial support during the completion of this work, within the program “New Developments in AdS3/CFT2 Holography”.

\begin{appendix}
\section{Explicit massless scalar modes}\label{appmodesols}
After the separation of variables \eqref{sepvar}, the Laplace equation \eqref{eomt} on AdS$_3$ is equivalent to the standard hypergeometric equation
\begin{equation}
z(1-z)\frac{d^2}{dz^2}g_{h,\bh}(z)+[c-(a+b+1)z]\frac{d}{dz}g_{h,\bh}(z)-ab\, g_{h,\bh}(z)=0\label{hyp}
\end{equation}
with $c=0$, $a=h$, $b=-\bh$ and the definitions
\begin{equation}
f_{h,\bh}(y)=\left(\frac{4-y}{4+y}\right)^{h-\bh}g_{h,\bh}(z)\qquad\qquad\qquad z=\frac{16y}{(4+y)^2}
\end{equation}
The asymptotic solutions \eqref{basisas} are thus extended to the bulk in terms of the solutions of this hypergeometric equation:
\begin{eqnarray}
t_{h,\bh}^+&=&\frac{16y\,e^{-i(hx_++\bh x_-)}}{(4+y)^2}\left(\frac{4-y}{4+y}\right)^{h-\bh}\,{}_2F_1(1+h,1-\bh,2;\frac{16y}{(4+y)^2})\\
t_{0,\bh,\sigma}^-&=&\frac{1-\s}{2}e^{-i\bh x_-}\left(\frac{4-y}{4+y}\right)^{\bh}+\frac{1+\s}{2}e^{-i\bh x_-}\left(\frac{4+y}{4-y}\right)^{\bh}\\
t_{h,0,\sigma}^-&=&\frac{1-\s}{2}e^{-ih x_+}\left(\frac{4-y}{4+y}\right)^{h}+\frac{1+\s}{2}e^{-ih x_+}\left(\frac{4+y}{4-y}\right)^{h}\\
t_{h,\bh}^-&=&e^{-i(hx_++\bh x_-)}\left(\frac{4-y}{4+y}\right)^{h-\bh}g^-_{h,\bh}(z)+a_{h,\bh}\, t^+_{h,\bh} \label{tminfull}
\end{eqnarray}
\begin{eqnarray*}
g^-_{h,\bh}(z)&=&\frac{\Gamma(1+h)\Gamma(1-\bh)}{\Gamma(1+h-\bh)}z\,{}_2F_1(1+h,1-\bh,1+h-\bh;1-z)\\
a^-_{h,\bh}&=&h\bh(2\gamma-1)\hspace{8cm}\mbox{when }h,-\bh >0\\
g^-_{h,\bh}(z)&=&\frac{\Gamma(1-h)\Gamma(1+\bh)}{\Gamma(1-h+\bh)}z(1-z)^{\bh-h}\,{}_2F_1(1-h,1+\bh,1-h+\bh;1-z)\\
a^-_{h,\bh}&=&\frac{\bh-h}{2}+h\bh(2\gamma-1)\hspace{6.6cm}\mbox{when }-h,\bh >0\\
g^-_{h,\bh}(z)&=&\frac{(-1)^h\Gamma(1+h)\Gamma(1+\bh)}{\Gamma(1+h+\bh)}z^{-h}\,{}_2F_1(h,1+h,1+h+\bh;z^{-1})\\
a^-_{h,\bh}&=&\frac{\bh-h}{2}+h\bh(2\gamma-1-i\pi)\hspace{5.8cm}\mbox{when }h,\bh >0\\
g^-_{h,\bh}(z)&=&\frac{\Gamma(1-h)\Gamma(1-\bh)}{(-1)^\bh\Gamma(1-h-\bh)}z^\bh\,{}_2F_1(-\bh,1-\bh,1-h-\bh;z^{-1})\\
a^-_{h,\bh}&=&\frac{\bh-h}{2}+h\bh(2\gamma-1-i\pi)\hspace{5.8cm}\mbox{when }-h,-\bh >0
\end{eqnarray*}
Note that the $g_{h,\bh}^-$ are analytic series around small $1-z$ or $z^{-1}$, not around $z=0$ which is the boundary. This way of presenting those functions is more compact, and one can recover the asymptotic expansion around $z=0$ via a standard reference on hypergeometric functions, e.g. \cite{specialfunc}, or a suitable computer algebra system. Those near boundary expansions will have the form of a product of $\log z$ with a hypergeometric function in $z$, plus some additional power series in $z$. The small $z$ expansion can then be translated to the small $y$ expansion. The leading terms of small $y$ expansion of the precise linear combination appearing in \eqref{tminfull} coincide with those given in \eqref{basisas}.

\section{Details on holographic renormalization}\label{appholren}
In this Appendix we provide more details on the holographic renormalization procedure for the gravity- scalar system.
We will simultaneously treat the cases of a free complex scalar and an interacting axion-dilaton, and will also include a U(1) Chern-Simons field
which is present in the black hole deconstruction setting. The latter does not couple to the other fields in the bulk, but can influence the asymptotic charges
as is known from the holographic realization of spectral flow \cite{David:1999zb},\cite{Maldacena:2000dr}.

 We start from the action for a complex scalar minimally coupled to   3D gravity with negative cosmological constant and a  U(1) Chern-Simons field:
\be
S = \int_\calm \left[  d^3 x \sqrt{- G} \left( \calr + {2 \over l^2} - {\pa_\m \t \pa^\m \bar \t\over 2 \a^2 (\Im \t)^2} \right)+{l\over 2} \cala \wedge d \cala\right]
 - 2 \int_{\pa \calm} \sqrt{-\g} K .\label{Sbare}
\ee
The constant $\a$ lets us interpolate between an axion-dilaton field for $\a = 1$  and a free scalar, which is obtained upon taking the limit
\be
\a \to 0 \qquad {\rm with\ } t =\t-{i \over \a} {\rm fixed}.\label{tofree}
\ee

The equations of motion following from (\ref{Sbare}) are
\be
\Box \t + i {\pa_\m \t \pa^\m  \t\over \Im \t}=0, \qquad
\calr_{\m\n} + {2 \over l^2} G_{\m\n} -{\pa_{(\m} \t \pa_{\n)} \bar \t\over 2 \a^2 (\Im \t)^2}=0, \qquad d \cala =0.
\label{eom3D}
\ee
We use Fefferman-Graham coordinates in terms of which the metric looks like
\be\label{FGcoords}
ds^2_3 = l^2 \left( {dy^2 \over 4 y^2} + {1 \over y} g_{ij} (x^k,y) dx^i dx^j\right)
\ee
The near-boundary expansion of the fields is then \cite{de Haro:2000xn}
\bea
t &=& t_{0} + y\, \t_{2} + y\log y\, \tilde t_{2}+ \calo ( y^2 \log y)\\
g_{ij} &=& g_{0\, ij}+ y\, g_{2\, ij}+ y\log y\, \tilde g_{2\, ij} + \calo ( y^2 \log y)\\
\cala &=& \cala_0 + \calo(y),\label{FGexpapp}
\eea
where the coefficient functions on the RHS are independent of $y$.
Substituting these in the equations  of motion (\ref{eom3D}) and working out the leading terms one finds that the logarithmic coefficients
$\tilde g_{2}, \tilde \t_{2}$ are completely determined  by the boundary values $ g_{0},  \t_{0}$:
\bea
\tilde \t_{2} &=& -{1\over 4} \Box_{0} \t_{0} - {i \over 4} {\pa_i \t_{0} \pa^i  \t_{0}\over \Im \t_{0}}\\
\tilde g_{2\, ij} &=& - {1  \over 4 \a^2 (\Im \t_{0} )^2}\left( {\pa_{(i} \t_{0} \pa_{j)} \bar \t_{0}}
-\half {\pa_k \t_{0} \pa^k \bar \t_{0}}g_{0\, ij}\right) \label{eomnbapp}
\eea
where indices are raised and covariant derivatives taken with respect to the boundary metric $g_{0}$. We note that $\tilde g_{2\, ij}$ is traceless.

For the tensor
 $g_{2\, ij}$  on the other hand, only the trace and divergence are fixed in terms of $ g_{0},   \t_{0}$:
 \be
 g_{2} = - \half R_{0} +   {\pa_i \t_{0} \pa^i \bar \t_{0} \over 4 \a^2 (\Im \t_{0} )^2}, \qquad
 \nabla^j ( g_{2ij}- g_{2} g_{0ij}) = {\Re\left(  \bar \t_{2} \pa_i \t_{0}  \right) \over \a^2 (\Im \t_{0} )^2}\label{g2solapp}
 \ee

  Proceeding as in \cite{de Haro:2000xn}, we regularize the action by cutting off the $y$ integral at $y= \e\ll 1$.
  One finds for the regularized on-shell action
  \be
  S_{\mathrm{reg}} = - {2 l} \int d^2 x \left[  \int_\e dy{\sqrt{-g }\over y^2}  + 2\left.\left( \pa_y\sqrt{-g } - {\sqrt{-g }\over y}\right)\right|_{y=\e} \right].\label{Sreg}
  \ee
  Using (\ref{FGexpapp}) one derives that this contains the following divergent terms as $\e \to 0 $:
  \be
  S_{\mathrm{div}} = l \int_{\pa \calm_\e}  d^2x\sqrt{- g_{(0)}}\left[  {2 \over \e }+ g_{(2)} \log \e \right]\label{Sdiv2}
  \ee
  We propose to add  the following additional boundary terms to the action:
\begin{align}
  S_{\mathrm{bnd}} = l \int_{\pa \calm_\e}  d^2x\sqrt{- g}&\left[ - {2 \over \e }+ \half \left( R  - {\pa_i \t \pa^i \bar \t\over 2 \a^2 (\Im \t)^2} \right)\log \e \right. \nonu
& \left. - {\Re (v^i \pa_i \t)}\left({1\over \a^2  \Im \t} -{1\over \a} + \a\right)+ w_{ij} g^{ij} \right].
  \end{align}
  The role of first line is to cancel the divergences of (\ref{Sdiv2}), while the second line serves to get a good variational principle under the boundary conditions to be specified below. It breaks boundary covariance and depends on a fixed, symmetric, lower index two-tensor $w_{ij}$ and a vector $v^i$ which will be specified below. Note that, unlike in the
  treatment of e.g. \cite{Kraus:2006nb}, we have not added any boundary terms for the Chern-Simons field $\cala$.

The variation of the total action then reads, up to bulk terms proportional to the equations of motion (\ref{eom3D}),
\begin{align}
\d( S+S_{\mathrm{bnd}}) =&  {l }\int_{\pa \calm_\e}  d^2x\sqrt{- g_{0}}\left[  \left(- g_{2\, ij}- \tilde g_{2\, ij} + g_{2}g_{0\, ij}  \right.\right.\nonu
& \left. + \half \left({1\over \a^2  \Im \t_0} -{1\over \a} + \a\right) \Re\left(  v^k\pa_k \t_{0}\right) g_{0\, ij} + w_{ij} -\half w  g_{0\, ij}\right)\d  g_{0}^{ij}\nonu
&\left. +{2\over \a^2 ( \Im \t_{0} )^2}\Re \left( (\t_{2}+ {i\over 2} v^i\pa_i \t_{0}+\tilde \t_{2})\d \bar \t_{0}\right)\right.\nonu
&\left. + {1 \over 2 \sqrt{-g_0}} \left( \cala_{0+} \d \cala_{0-} - \cala_{0-} \d \cala_{0+}\right) \right]\label{varS}
 \end{align}
Guided by    \cite{Compere:2013bya} for the choice of $w_{ij}$ and the choice (\ref{bdytermAdS}) for $v^i$ in a fixed background, we will take
\be
v^i =\d^i_-, \qquad w_{ij} =  {\D \over k} \d_i^+ \d_j^+.
\ee
We now propose our boundary conditions:
 \begin{align}
\t_{0} &= t_{0} (x_-), & \t_{2} &=-{i\over 2} \t_{0}'\label{CRBmetr1app}\\
 g_{0\,--} &= \bar P'(x_-), & g_{2\,++} + {1 \over 4} \cala_{0+}^2 &= {\D \over k} \\
  g_{0\,+-} &= -\half, & g_{0\,++} &= 0\\
  \cala_{0+} &= {\rm constant}, & \cala_{0-} &= \cala_{0+}  \bar P'(x_-).\label{CRBmetr4app}
\end{align}
 where $k \equiv 4 \p l$ and  $\D$ is a fixed constant  which specifies the theory, i.e. $\d \D=0$.  We note that
 the value of $\D$ changes if we turn on the Chern-Simons field $\cala$. This is analogous to the  bulk realization of spectral flow \cite{David:1999zb},\cite{Maldacena:2000dr}
 in the case of conformal boundary conditions.

 Furthermore, the equations  following  from (\ref{eomnbapp})  imply
 \begin{align}
 \tilde t_{2} &=0, &
 \tilde g_{2\, ij} &= - {|\t_{0}'|^2\over 4 \a^2 (\Im \t_0)^2}\d_i^- \d_j^- \\
  g_{2+-} &= -{\D \over k} \bar P' + {1\over 4} \cala_{0+}^2,&
  \pa_+   g_{2\,--} &= 0
 \end{align}
 These boundary conditions (\ref{CRBmetr1app}-\ref{CRBmetr4app}) lead to a good variational principle, as one can check that plugging  into (\ref{varS}) one obtains $\d( S+S_{\mathrm{bnd}})=0$.

They are also invariant under the asymptotic symmetries generated by (\ref{asU1},\ref{asvir}), with the free functions transforming as
 \begin{align}
    \d_{\bar V} \bar P'  &= \pa_- ( \bar V\bar P')  & \d_{\bar U} \bar P'  &= - \bar U'  \\
     \d_{\bar V} \bar L  &= \bar V \bar L' + 2 \bar V' \bar L - {k\over 2} \bar V''' - {k\over 4} \bar V' {|\t_{0}'|^2 \over \a^2 (\Im \t_0)^2}& \d_{\bar U}\bar L  &= 0\\
      \d_{\bar V} \t_{0} &= \bar V  \t_{0}'  & \d_{\bar U} \t_{0} &= 0
   \end{align}

\section{First order solution}\label{apppert}
In this Appendix we construct the  the backreacted metric of a collection point particles on helical geodesics, to first order in an expansion in the (assumed small) mass parameters $m_i$.
From (\ref{Liouv}) we find that the  first order correction of the Liouvile field should satisfy
\be
\left( \pa_v \pa_{\bar v} -2 e^{-2 \F_{\rm AdS}}\right) \F_1 = {1 \over 4}\sum_i {m_i} \d^2 (v-v_i, \bar v - \bar v_i),\label{Phi1eq}
\ee
where $\F_{\rm AdS}$ is the global AdS$_3$ solution (\ref{PhiAdS}).
One way to solve this equation would be to use a Green's function but we will use here a  simpler method using conformal mapping. We start by constructing the solution for $\F_1$ corresponding to a single mass $m$ in the origin $z=0$. As explained in \cite{Raeymaekers:2015sba}, this can be found by using rotational symmetry to reduce the problem to a second order ordinary differential equation, and to fix the integration constants to get the correct delta-function source in (\ref{Phi1eq}) and the desired near-boundary  behaviour $\F_1 \to 0$ for $|v|\to 1$. This leads to
\be
\F_1={m \over 4 \p}\left( 1 + { 1 + |v|^2 \over 1- |v|^2} \ln |v| \right).\label{Phicenter}
\ee
To find the solution corresponding to a point mass away from the origin, say in $v=v_1$,  we apply  the M\"{o}bius transformation which maps the origin to $v_1$:
\be
v \to w={v - v_1 \over 1 - v \bar v_1}.\label{Mobius}
\ee
The AdS solution $\F_{\rm AdS}$ is essentially the conformal factor in front of  the metric on the Poincar\'e disk, $e^{-2 \F_{\rm AdS} }d z d\bar z$.
The M\"{o}bius transformation (\ref{Mobius}) is an isometry of this metric and hence leave  it form-invariant, which amounts to the following $\F$-transformation
\be
e^{ - 2 \F_{\rm AdS} ( w )} =\left| {\pa v \over \pa w} \right|^2 e^{ - 2 \F_{\rm AdS} ( v )}
\ee
as one can easily check from the explicit expression (\ref{PhiAdS}). Substituting this in (\ref{Phi1eq}), we see that the solution for $\F_1$
for  a point mass in $v=v_1$ can be simply obtained from (\ref{Phicenter}) by applying the  M\"{o}bius transformation (\ref{Mobius}). Using the linearity
of (\ref{Phi1eq}) we then obtain the first order solution $\F_1$ for  multiple point masses:
\be
\F_1 ={1 \over 4 \p} \sum_i m_i \left( 1 - { |v-v_i|^2 + |1 - \bar v_i v|^2\over |v-v_i|^2 - |1 - \bar v_i v|^2}\ln {|v-v_i|\over  |1 - \bar v_i v|} \right).\label{1storderapp}
\ee

\end{appendix}

\end{document}